\documentclass[fleqn,usenatbib]{mnras}
\usepackage{newtxtext,newtxmath}
\usepackage[T1]{fontenc}
\usepackage{amsmath}
\usepackage{multirow}
\usepackage{multicol}
\usepackage{graphicx}
\usepackage{hyperref}
\usepackage{cleveref}
\usepackage{threeparttable}
\usepackage[dvipsnames]{xcolor}
\usepackage{ulem}
\usepackage{orcidlink}
\usepackage{subcaption}
\usepackage[percent]{overpic}

\graphicspath{{./}{figures/}}

\newcommand{\nodata}{\dots}
\newcommand{\iwd}{\,(\checkmark)}

\title[Scattering toward the near Galactic bar end]{Heavy interstellar scattering toward the near end of the Galactic bar}

\author[Pushkarev et al.]{A.~B.~Pushkarev \orcidlink{https://orcid.org/0000-0002-9702-2307},$^1$\thanks{E-mail: pushkarev.alexander@gmail.com}
A.~Brunthaler \orcidlink{https://orcid.org/0000-0003-4468-761X},$^2$
Y.~Y.~Kovalev \orcidlink{https://orcid.org/0000-0001-9303-3263},$^2$
M.~M.~Lisakov \orcidlink{https://orcid.org/0000-0001-6088-3819},$^3$
I.~N.~Pashchenko \orcidlink{https://orcid.org/0000-0002-9404-7023},$^{4,5}$
\newauthor
A.~V.~Plavin \orcidlink{https://orcid.org/0000-0003-2914-8554},$^6$
N.~Roy \orcidlink{0000-0001-9829-7727},$^{7,8}$ 
P.~A.~Voitsik \orcidlink{https://orcid.org/0000-0002-1290-1629},$^4$
S.~A.~Dzib \orcidlink{https://orcid.org/0000-0001-6010-6200},$^2$
T.~A.~Koryukova \orcidlink{https://orcid.org/0000-0001-8347-7880},$^{4,1}$
A.~Y.~Yang \orcidlink{0000-0003-4546-2623}$^{9,10}$\\
$^1$Crimean Astrophysical Observatory, 298409 Nauchny, Crimea\\
$^2$Max-Planck-Institut f\"ur Radioastronomie, Auf dem H\"ugel 69, D-53121 Bonn, Germany\\
$^3$Instituto de Física, Pontificia Universidad Católica de Valparaíso, Casilla 4059, Valparaíso, Chile\\
$^4$Lebedev Physical Institute, Profsoyuznaya 84/32, Moscow 117997, Russia\\
$^5$Moscow Institute of Physics and Technology, Institutsky per. 9, Dolgoprudny, Moscow region, 141700, Russia\\
$^6$Black Hole Initiative at Harvard University, 20 Garden Street, Cambridge, MA 02138, USA\\
$^7$Department of Physics, Indian Institute of Science, Bengaluru 560012, India\\
$^8$Department of Physics, New Mexico Institute of Mining and Technology, Socorro, NM 87801, USA\\
$^9$National Astronomical Observatories, Chinese Academy of Sciences, Beijing 100101, China\\
$^{10}$Key Laboratory of Radio Astronomy and Technology, Chinese Academy of Sciences, A20 Datun Road, Chaoyang District, Beijing, 100101, China
}

\date{Accepted 2026 February 4. Received 2026 February 4; in original form 2025 December 13}

\pubyear{2026}

\begin{document}
\label{firstpage}
\pagerange{\pageref{firstpage}--\pageref{lastpage}}
\maketitle
    
\begin{abstract}
We present results of a pilot observational wide-field VLBI campaign on probing scattering properties of the partly ionized interstellar medium towards the Galactic plane sky region between $28^\circ<l<36^\circ$ and $|b|<1^\circ$. This covers the region where the Galactic bar connects to the spiral arms and where a lot of star formation is currently ongoing. The Very Long Baseline Array (VLBA) observations of the whole region were performed in a special mode with multiple phase centers at L-band (1.4 -- 1.8~GHz) during April-June 2022 and a year later complemented by sessions at S (2.2 -- 2.4~GHz) and C-band (4.6 -- 5.0~GHz) partially covering the pilot region. We found compelling evidence that target sources are subject to scattering. The total detection rate in L, S and C-bands is 1.5, 3.4 and 9.2 per cent, respectively, and approximately scales with the square of the observation frequency. The low rate values imply that scattering is strong. Its power is non-uniform across the Galactic plane and it can be approximated by a Gaussian with a width of about $2^\circ$ peaking at the Galactic mid-plane. One of the brightest sources of the field shows anisotropic scattering, with a $\lambda^2$ dependence of its observed angular size, along a position angle of $26^\circ$ aligned with the line of constant Galactic latitude. We estimate the turbulence dissipation scale $r_\text{in}\approx1500$~km toward the source J1833+0015.
\end{abstract}
    

\section{Introduction}

The interstellar medium (ISM) of the Milky Way is complex, highly inhomogeneous, and often turbulent due to a wide variety of energetic processes originating from supernova explosions, stellar winds, differential rotation of the Galaxy, large-scale gravitational instabilities, and density waves developing in the spiral arms \citep{Low04,Elmegreen04,Piontek05,Krumholz16}. The corresponding energetic cascades generate turbulence, sustain it, and transform the energy through non-linear dynamics to progressively smaller spatial scales until it is damped by viscous dissipation. Radio emission passing through the thermal plasma of the ISM containing spatial fluctuations of density of free electrons $\delta n_\text{e}/n_\text{e}$ is subject to scattering effects. The diversity of the propagation effects comes from the parameters of an assumed three-dimensional power spectrum of turbulence \citep{Blandford85,Rickett90}:
\begin{gather}
P_{\delta n_\text{e}} (q) = C_n^2 q^{-\beta}\,, \qquad q_0\lesssim q\lesssim q_1\,,
\end{gather}
\noindent
where $C_n^2$ is a normalizing constant depending on the scattering strength, and $q_0$ and $q_1$ are the spatial wave-number cutoffs that correspond to the outer $r_\text{out}=2\pi/q_0$ and inner $r_\text{in}=2\pi/q_1$ scales of turbulence, respectively, covering many orders of magnitude, from $10^8$ to $10^{20}$~cm in the ionized ISM \citep{Sprangler90,Armstrong95,Combes00,Chepurnov10,Johnson18}. For a Kolmogorov spectrum of density fluctuations, $\beta=11/3$ \citep{Goldreich95}.

However, in real astrophysical conditions the ISM in the Galaxy consists of well-separated clumps, and in most of the known cases only one of these clumps dominates the scattering on each line of sight. In these circumstances, it is possible to approximate scattering effects as they originate from a thin two-dimensional screen. Turbulent screens with more shallow spectra ($\beta<3.7$), characterized by predominantly small-scale inhomogeneities, produce diffractive-dominated scattering effects: (i) angular broadening of compact background radio sources, e.g. pulsars, masers, active galactic nuclei (AGN) \citep[e.g.,][]{Fey89,Lazio04}, (ii) relatively fast flux modulations (hours-days), known as scintillations \citep[e.g.,][]{Savolainen08,Koay19}. On the contrary, for screens with steeper spectra ($\beta>3.7$) and thus having large-scale density fluctuations or discrete structures, scattering manifests itself through refractive dominated effects: (i) multiple imaging detected as parabolic arcs in secondary dynamic spectra of pulsar observations \citep{Cordes86} or direct detection of secondary images of AGN cores induced by scattering \citep{Pushkarev13,Koryukova23}, (ii) angular position wander \citep{Clegg98}, (iii) relatively slow symmetric flux variations at scales from weeks to years with formation of caustic surfaces, the so-called extreme scattering events \citep{Fiedler87,Maitia03,Koryukova25}, associated with plasma lens crossing the line of sight to the background source. However, the presence of parabolic arcs does not necessarily imply a steep spectrum of turbulence \citep{Tuntsov13}.

\begin{figure*}
\centering
\includegraphics[width=\linewidth]{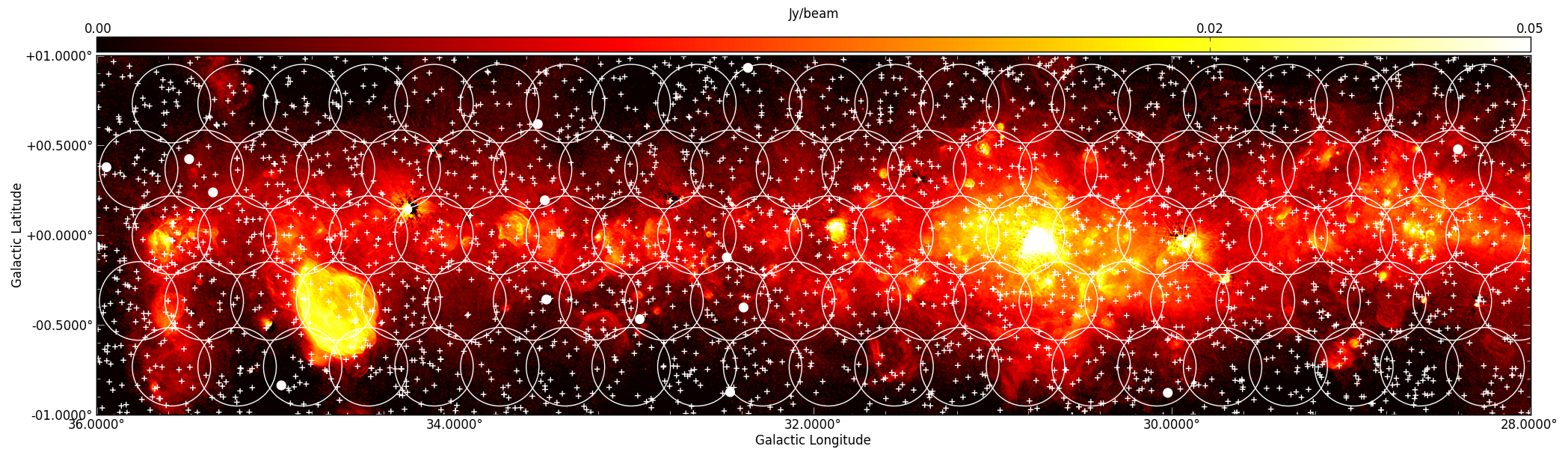}
\caption{The full $8^\circ\times2^\circ$ image of the pilot region from the combination of VLA D-configuration and Effelsberg single-dish data \citep{Brunthaler21} together with the proposed VLBA pointing positions. Shown are pointing locations with the primary beam of the VLBA at 1.6 GHz (white circles), the location of already known VLBI sources (white dots), and all compact sources discovered in the VLA B-configuration images (white crosses).}
\label{f:pilot}
\end{figure*}

Scattering is most effective within the ISM of the Galaxy. Angular broadening induced by the ionized intergalactic medium (IGM) is negligible and can barely be detected by ground-based interferometers. Simulations in \cite{McQuinn14} have shown that the emission of an extragalatic radio source at a cosmological distance ($z\gtrsim0.5$) likely passes through regions within a virial radius of a galaxy halo of $10^{10}\text{M}_\mathrm{\sun}$. Even under this condition, the IGM scatter broadening is less than about 3~$\mu$as at 5~GHz \citep{Koay15}.

Within the Galaxy, the scattering power of the interstellar medium exhibits significant spatial variations. Angular broadening toward low Galactic latitudes was originally detected by \citet{Duffett76}, who observed a sample of 32 quasars at 151~MHz. This effect was later confirmed using a much larger sample of about 3000 AGN, which showed an excess in their angular sizes if seen through the Galactic plane ($|b|\lesssim10^\circ$) at $\nu\lesssim8$~GHz \citep{Pushkarev15}. Expanding the sample further to about 9000 sources enabled \citet{Koryukova22} to construct the first all-sky map of scattering power, revealing pronounced enhancements toward spiral arms, supernova remnants (SNR), and the Galactic center, while inter-arm regions exhibit much weaker scattering \citep{Fey91}. On smaller angular scales -- within a few degrees of the Galactic plane in the Cygnus region -- the scattering measure varies substantially, indicating that the ISM is highly clumpy \citep{Fey89}. The strongest scattering is observed toward the Galactic center, Sgr~A*, the angular size of which scales as $1.38\lambda_\text{cm}^2$~mas along the major axis \citep{Johnson18}, corresponding to roughly 500~mas at 1.58~GHz.

 The GLObal view on STAR formation (GLOSTAR) survey \citep{Medina19, Brunthaler21} has led to the detections of thousands of compact radio continuum sources in the Galactic plane \citep{Dzib23, Yang23}. GLOSTAR covers the target region using the VLA C-band (4.2 -- 5.2~GHz and 6.4 -- 7.4~GHz) in B- and D-configuration (angular resolution of $\sim$1.5$^{\prime\prime}$ and $\sim$18$^{\prime\prime}$, respectively), with zero spacings from the Effelsberg telescope covering 4 -- 8~GHz. Based on multi-wavelength data it was possible to classify the majority of the sources into galactic source candidates (e.g. H\,{\sc ii} regions, planetary nebulae, radio stars) and extragalactic source candidates. There are several types of galactic radio sources that could contain very compact radio emission on milliarcsecond scale. These include pulsars, black hole X-ray binaries, radio stars, or compact synchrotron jet from young stars. Furthermore, many extragalactic radio sources have very compact emission from an AGN. While extragalactic sources should be stationary on the sky, the proper motions of sources in the Milky Way are detectable with VLBI observations within one year. With the observations described in this paper, we aimed at detecting compact structure on VLBI scales and find new black hole X-ray binaries, detect new radio stars, and study the scattering properties of the ISM in the Milky Way with an unprecedented spatial resolution. In this paper, we study scattering properties of the ISM toward a region where the Galactic bar ends and connects to the spiral arms.

\section{Observations and data processing}

We observed almost the entire 16 square degrees of the pilot region of the GLOSTAR survey, covering the range from $l=28^\circ$ to $36^\circ$ and $|b|<1^\circ$ with the VLBA at L-band (1.4 -- 1.8~GHz). The field was covered by 107 pointings with the VLBA using a hexagonal grid with a grid spacing of $22^\prime$, corresponding to $1/\sqrt{2}$ of the primary beam at 1.4~GHz (see \autoref{f:pilot}). The 107 pointings were spread over 6 observations, that were conducted between 2022 April 10 and 2022 June 13. Each observation session lasted 6 hours, and each pointing was observed 5 times for 155 seconds to improve the uv-coverage, giving us $\sim$13 minutes integration time per pointing. We used 1830+012 (J1833+0115) as a phase calibrator separated by a few degrees from the target objects, and the ICRF \& RFC source 2007+777 \citep{Charlot20,RFC} as a fringe finder. Additionally, we observed six known VLBA calibrators near our target region to check whether they could be used as potential phase-reference. Since they are located at different distances from the Galactic plane, they can also be used to investigate the scattering properties off the plane. The observations were conducted with a total recording rate of 2048~Mbps, with eight 32-MHz-wide intermediate frequency channels (IF), with 64 spectral channels per IF, in dual polarization mode using 2-bit sampling. The integration time per source was chosen to reach a theoretical noise of about 70~$\mu$Jy in the phase-referenced images, which roughly matches the detection threshold of the GLOSTAR observations. The data were correlated with the DiFX correlator \citep{Deller2007} in Socorro, New Mexico, using an averaging time of 1~sec. These pointings covered 1210 compact sources from the GLOSTAR catalog, and we used the multi-phase center capability of the DiFX correlator.

\begin{figure*}
\centering
\includegraphics[width=\linewidth]{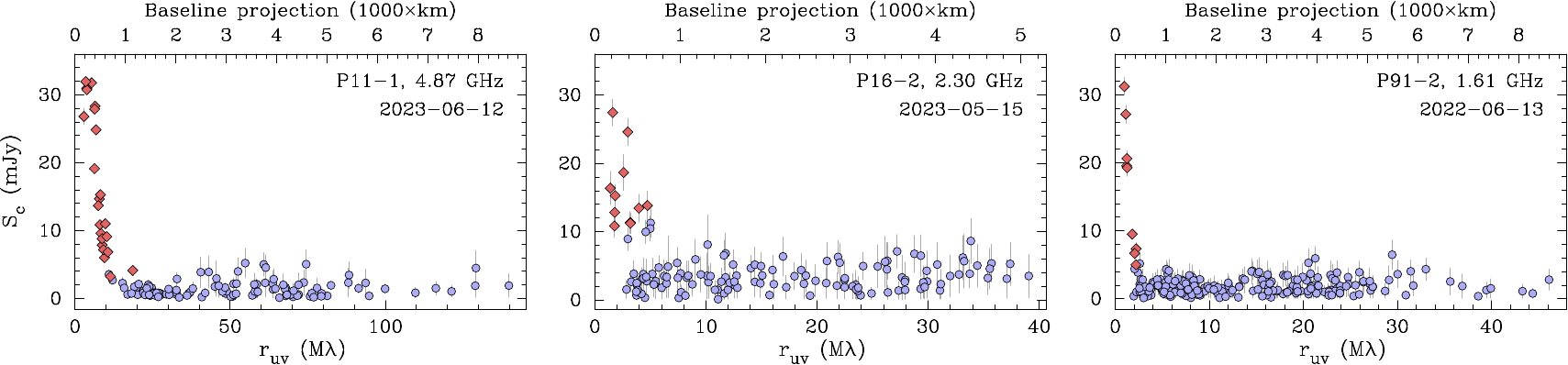}
\caption{Correlated flux density, $S_\text{c}$, as a function of baseline projection $r_{uv}$ for one of the target sources, 1847+007, detected at three frequency bands, C (left), S (middle) and L (right). Every point marks a vector average over a scan length of $\sim$2.5~min and over IFs on an individual interferometer baseline. Red diamonds represent visibilities with $\text{S/N}>6$, while lavender points show the measurements with a lower significance. Data were phase-calibrated to a calibrator. Error bars represent statistical $1\sigma$ uncertainties. Each plot legend shows the observing epoch, frequency and a source name in the internal notation PXX-X (note, sources have different internal names at different frequency bands, see \autoref{t:detected_sources}). Rapid decrease of the visibility function amplitudes with baseline length indicates that the source is heavily resolved. The plots for all detected sources are shown in \autoref{fa:radplots_c}, \autoref{fa:radplots_l}, and \autoref{fa:radplots_s}.}
\label{f:radplots}
\end{figure*}

The results we have obtained from the initial analysis at L-band (Sec.~\ref{s:detection_rate}) showed that the vast majority of target sources were not detected. It became clear that the scattering power toward the pilot sky region is much stronger than we expected. For this reason, the rest of the project time ($\sim$45 per cent) was used for observations at higher frequencies: one session at S-band (2.3~GHz, covering $\sim$8 per cent of the pilot region) and four sessions at C-band (4.8~GHz, covering $\sim$24 per cent), with 87 and 185 sources, respectively. We selected the brightest objects and performed targeted observations on them, plus sources that were in the primary beams of these. Each of these segments lasted 6 hours. The S-band observations were performed on May 15, 2023 using eight 32~MHz-wide IFs with 64 spectral channels per IF. The C-band sessions were conducted between June 12 and July 4, 2023 having four 128~MHz-wide IFs with 64 spectral channels per IF. Each target source had five and four 155-sec scans at S and C-band, respectively, while each scan of the phase calibrator was $\sim$1~min at both bands. The frequency setups and session dates for all three bands are listed in \autoref{ta:freqs} and \autoref{ta:sessions}.

For all frequency bands, we reduced the data with the NRAO Astronomical Processing System \citep[{\sc aips},][]{AIPS} following the standard procedure. To calibrate amplitudes, the antenna gain curves and system temperatures routinly measured during the observing sessions were applied. To improve data calibration, we performed task {\sc bpass} to determine the complex bandpass response function. Global gain corrections were derived for each station and IF from self-calibration of the phase calibrator J1833+0115 and applied using task {\sc clcor}. As most of the target sources are weak, the observations were performed in a phase reference mode with the nodding style \citep{Beasley95}, alternating between a target source and a nearby relatively bright ($\sim$300~mJy) phase calibrator. This allows to accurately transfer phases from a calibrator to a target source by interpolating these phases to the interleaved target scans. For global fringe-fitting with phase referencing, the procedure {\sc vlbafrgp} was run. To apply calibration for the secondary phase centers, we used the {\sc vlbamphc} pipeline. Throughout data reduction, different IFs in each frequency band were kept separately and processed individually. To prevent generating a weak fake signal (an additional delta-function component) from noise, we omitted the initial phase-only self-calibration using a point-source model (task {\sc calib}). Instead, the calibrated data were saved on disk using task {\sc split}. As a result, the target sources remain offset from the phase center, due to the inherent astrometric uncertainty within $0.1^{\prime\prime}$ in their coordinates.

\section{Results}

\subsection{Detection rate}
\label{s:detection_rate}

After applying the calibration of the phase calibrator, we imaged all 1210 sources in AIP2021S using natural weighting. The size of the individual images was $0.5^{\prime\prime}$ and much larger than the position uncertainty from our GLOSTAR VLA observations. Even though we reached close to the thermal noise limit in the images from most of the sources, we were able to reveal only a single clear detection (see Section~\ref{s:G_detection}). However, there were a number of sources that had an increased noise in the images. After inspecting the uv-data of these sources, it became clear that they were only detected on the shortest baselines of the VLBA and completely resolved out on longer baselines. We therefore inspected the visibilities for all the targets to detect more sources.

\begin{figure*}
\centering
\includegraphics[width=0.75\linewidth,trim={0 22 0 0},clip]{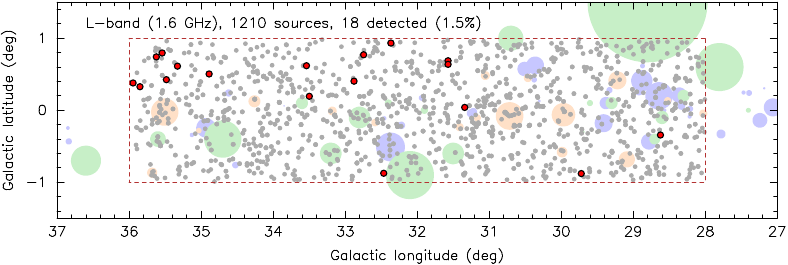}
\includegraphics[width=0.75\linewidth,trim={0 22 0 0},clip]{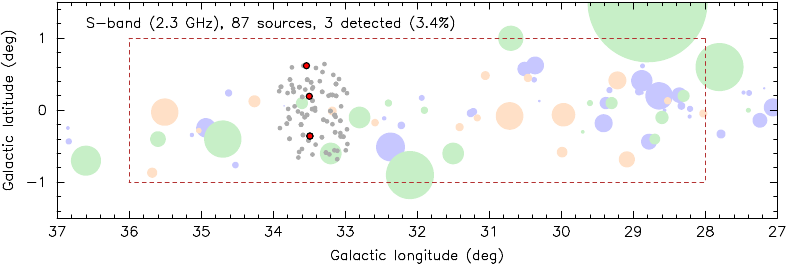}
\includegraphics[width=0.75\linewidth]{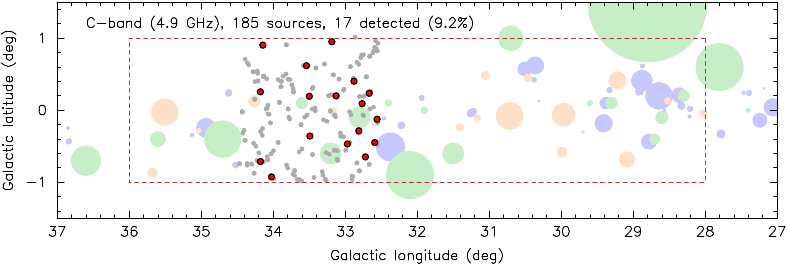}
\caption{Positions of the observed sources in the pilot region. Grey points show undetected objects. Red points represent detected sources. Light-green circles depict positions and sizes of 20 known SNR within the sky area \citep{Green25}, blueish ones represent 42 SNR candidates from the GLOSTAR survey \citep{Dokara21}, and 17 peach circles denote H\,{\sc ii} regions \citep{Medina19}. Detection rate increases with frequency as scatter broadening weakens.}
\label{f:sky_coverage}
\end{figure*}

\begin{table*}
    \centering
    \caption{Detected sources in the presented VLBA observations.}
    \label{t:detected_sources}
    \begin{tabular}{cccccccc}
    \hline\hline
GLOSTAR & Other  &\multicolumn{3}{c}{VLBA label} &         R.A.                   & Dec.  & Obj. \\ \cline{3-5}
Name    & Name   & 4.87\,GHz  & 2.30\,GHz & 1.61\,GHz & ($^{\rm h}:^{\rm m}:^{\rm s}$) & ($^{\circ}:':"$)      & Type \\
\hline
G028.6204$-$00.3437 & GPSR5 28.619$-$0.343             & \nodata   & \nodata   & P6-2\iwd    &18:44:47.28393 &$-$03:59:36.5462 & EgC\\
G029.7196$-$00.8789 & [BHW2010] G29.7195$-$0.8788      & \nodata   & \nodata   & P9-8\iwd    &18:48:42.42154 &$-$03:15:34.7250 & EgC\\
G031.3392+00.0392  & PSR B1845$-$01                   & \nodata   & \nodata   & P47-40\iwd  &18:48:23.58677 &$-$01:23:58.3785 & PSR\\
G031.5694+00.6870  & \nodata                         & \nodata   & \nodata   & P52-8\iwd   &18:46:30.39742 & $-$00:53:57.1031 & EgC\\
G031.5711+00.6365  & GPSR 031.571+0.637              & \nodata   & \nodata   & P52-9\iwd   &18:46:41.34912 &$-$00:55:14.5821 & EgC\\
G032.3644+00.9326 & [IBR2011] J1847$-$0004            & \nodata   & \nodata   & P79-16\iwd  &18:47:04.95956 &$-$00:04:47.1494 & EgC\\
G032.4643-00.8749 & [IBR2011] J1853$-$0048            & \nodata   & \nodata   & P58-1\iwd   &18:53:41.98072 &$-$00:48:54.3051 & EgC\\
G032.5569$-$00.1246 & GPSR5 32.557$-$0.125              & P33-1\iwd & \nodata   & P80-6       &18:51:11.83963 & $-$00:23:25.9597  & EgC\\
G032.5898$-$00.4469 & {[BHW2010]} G32.5898$-$0.4468     & P25-2\iwd & \nodata   & P59-17      &18:52:24.28653 & $-$00:30:29.5276  & EgC\\
G032.6673+00.2362  & ...                             & P23-1\iwd & \nodata   & P81-14      &18:50:06.85728 & $-$00:07:40.1887  & EgC\\
G032.7194$-$00.6478 & {[BHW2010]} G32.7193$-$0.6477     & P64-1\iwd & \nodata   & P58-6       &18:53:21.37676 & $-$00:29:04.3148  & EgC\\
G032.7440+00.7700 & NVSS J184821+001108              & P15-1     & \nodata   & P82-13\iwd  &18:48:21.22303 & +00:11:01.8494 & EgC\\
G032.7635+00.0915  & GPSR5 32.763+0.091              & P13-1\iwd & \nodata   & P80-18      & 18:50:48.30295 & $-$00:06:29.3853 & EgC \\
G032.8100$-$00.2862 & GPSR5 32.810$-$0.286              & P58-1\iwd & \nodata   & P61-9       &18:52:14.06579 & $-$00:14:20.5442  & EgC\\
G032.8776+00.4038  & GPSR5 32.878+0.404              & P19-1\iwd & \nodata   & P84-8\iwd   &18:49:54.08514  & +00:08:08.8555 & EgC\\
G032.9686$-$00.4681 & {[IBR2011]} J1853$-$0010          & P5-1\iwd  & P1-1      & P63-1       & 18:53:10.25911 & $-$00:10:50.7605 & EgC \\
G033.1860+00.9528  & ...                             & P73-1\iwd & \nodata   & \nodata     &18:48:30.53892 & +00:39:37.9663  & EgC\\
G033.4894$-$00.3585 & GPSR5 33.489$-$0.358              & P14-1\iwd & P12-2\iwd & P65-4       &18:53:43.85640  & +00:19:57.3973 & EgC\\
G033.4980+00.1943  & QSO J1851+005                   & P1-1\iwd  & P15-6\iwd & P90-5\iwd   & 18:51:46.72431 & +00:35:32.4499  & QSO \\
G033.5372+00.6191  & {[L2004b]} J185020.140+004913.91& P11-1\iwd & P16-2\iwd & P91-2\iwd   & 18:50:20.24981 & +00:49:15.3263  & EgC \\
G033.7039$-$00.9624 & ...                             & P133-1\iwd& \nodata   & P64-8       &18:56:16.33143 & +00:14:53.1113  & EgC\\
G034.0218$-$00.9255 & {[L2004b]} J185643.059+003250.93& P17-1\iwd & \nodata   & P66-4       &18:56:43.23155  & +00:32:52.3812 & EgC\\
G034.1428+00.9040  & ...                             & P94-1\iwd & \nodata   & P94-6       &18:50:25.64885 & +01:29:23.2806  & EgC\\
G034.1777$-$00.7115 & ...                             & P21-1\iwd & \nodata   & P66-13      &18:56:14.61089 & +00:47:03.0066  & EgC\\
G034.1782+00.2564  & ...                             & P43-1\iwd & \nodata   & P96-9       &18:52:47.89478 & +01:13:33.8027  & EgC\\
G034.8893+00.5042 & \nodata                          & \nodata   & \nodata   & P102-2\iwd  &18:53:12.80866 & +01:58:19.1629 & EgC\\
G035.3304+00.6133 & \nodata                          & \nodata   & \nodata   & P103-9\iwd  &18:53:37.78147 & +02:24:51.1876 & EgC\\
G035.4843+00.4239 & NVSS J185435+022753              & \nodata   & \nodata   & P105-12\iwd &18:54:35.11222 & +02:27:53.4398 & EgC \\
G035.5412+00.7951 & \nodata                          & \nodata   & \nodata   & P106-6\iwd  &18:53:21.99018 & +02:41:04.9072 & EgC\\
G035.6234+00.7413 & \nodata                          & \nodata   & \nodata   & P106-12\iwd &18:53:42.50671 & +02:44:00.0146 & EgC\\
G035.8491+00.3278 & GPSR5 35.849+0.328               & \nodata   & \nodata   & P107-11\iwd &18:55:35.63212 & +02:44:44.0340 & EgC\\
G035.9464+00.3787 & NVSS J185535+025119              & \nodata   & \nodata   & P107-13\iwd &18:55:35.43662 & +02:51:19.4369 & QSO\\
\hline
    \end{tabular}
{\bf Notes.} From left to right the columns are: GLOSTAR designation from \citet{Dzib23}, alternative source name, labels used in VLBA observations at the three observed bands (labels marked with \iwd\, indicate detections), R.A. and declination coordinates used for correlation, and object type. Object type follows \citet{Dzib23}: EgC = Extragalactic candidate, PSR = Pulsar, and QSO = quasar.
\end{table*}

To increase sensitivity of the visibilities, we combined all IFs together. To identify whether a source is detected or not, we visually inspected its visibility function (both amplitudes and phases). We treated a source as detected if it showed an increase in the correlated flux density and a better phase stability toward progressively smaller baseline projections. In 75 per cent of detections, the correlated flux density exceeded $6\sigma$, in the rest 25 per cent it was over the $4\sigma$ level. In \autoref{f:radplots}, we present the so-called radplots, i.e., amplitude of the visibility function versus projected spacing for the known VLBI source 1847+007 ($l=33\fdg537$, $b=0\fdg619$), one of two sources detected at three frequency bands. Results on the other source, the bright quasar 1849+005 ($l=33\fdg498$, $b=0\fdg194$), will be presented in a companion paper, as it shows indications of refractive substructure. We also note that in \cite{Plavin26} we present the first confident detection of refractive substructure in AGN with ground-based VLBI.

We detected 33 radio sources in our VLBA observations across the three observed frequency bands (4.87, 2.30, and 1.61 GHz). Table \ref{t:detected_sources} summarizes these detections and includes their GLOSTAR designations, VLBA labels used at each band, and positions adopted during data processing. The sample consists predominantly of sources classified as extragalactic candidates (EgC), which are likely quasars. As discussed by \citet{Dzib23}, this classification is primarily based on the absence of counterparts at other wavelengths, as well as on their radio emission properties, such as negative spectral indices and non-variability, characteristics that resemble those of background quasars. In addition, we also detected one known pulsar (see section~\ref{s:G_detection}) and two confirmed quasars. In \autoref{f:sky_coverage}, we show the distributions of the observed and detected sources. There are several features of the detected sources and their distributions at different frequencies:

(i) The number of detected sources is quite low.

(ii) The detection rate rapidly increases with frequency, 1.5 per cent at 1.6~GHz, 3.4 per cent at 2.3~GHz, and 9.2 per cent at 4.9~GHz. Fitting a detection rate as $\propto\nu^a$ with Binomial likelihood yields $a = 1.61\pm0.28$, which is significantly non-zero, but it is $<2$ that would be expected if source detectability is limited mainly by scattering broadening at all frequencies used in the fit. Thus, the fitted value of $a<2$ might indicate that at least some sources are not strongly affected by scattering at 5~GHz. However, the more likely reason is that we derive the detection rate statistics from the sky regions that are not equal across the different frequency bands. Additionally, a wide distribution of source flux densities and intrinsic spectrum can also affect the detection rate. Indeed, if we exclude three detected relatively bright VLBI sources (1849+005, 1847+007, and 1851+002), we obtain $a=1.96$. Finally, we note that the intrinsic frequency dependence of the detection rate for core-dominated AGN is itself complex. In the absence of scattering, the detection rate generally increases slightly with frequency. This trend is mainly driven by the higher compactness and flat or inverted spectra of the AGN cores, which make them brighter and more detectable at higher frequencies.

(iii) At the L-band, the detected sources tend to be located towards larger values of $|b|$. According to the Kolmogorov-Smirnov test, the distributions in galactic latitude between the detected and observed sources are significantly different, with $p=0.025$.

(iv) The detected sources tend to avoid the regions covered by SNRs. Indeed, none of the detected sources is seen through the SNR, where the ISM is highly turbulent, making source detection even more challenging. The most up-to-date sample of SNR compiled by \cite{Green25} comprises 310 objects. Their positions strongly concentrate towards the Galactic plane (91 per cent within $|b|<5^\circ$), and two-thirds of them are within $50^\circ$ the Galactic center. Our pilot sky region contains 18 supernova remnants \citep{Green25} and 32 SNR candidates \citep{Dokara21} altogether covering about 11 per cent of the area (\autoref{f:sky_coverage}, top panel). In fact, the overall SNR population is expected to be much more numerous and comprise about 2000 additional objects \citep{Ranasinghe21}, undiscovered due to their small angular sizes, low surface brightness, or misclassification with H\,{\sc ii} regions and planetary nebulae \citep{Ball23,Anderson25}. 
We also added 17 H\,{\sc ii} regions identified by \cite{Medina19} in the pilot sky area, including the well known massive star forming complex W43. These objects, expected to enhance scattering, cover about 3 per cent of the pilot region. None of the detected sources are seen through them, which means that the corresponding chance probability $p=(1-0.14)^{33}=0.007$.

All of the features listed above suggest that radio emission of the sources observed through the entire pilot sky region is subject to strong scattering. We note that a detection rate for the mJIVE-20 survey, which targeted more than 20k sources from the FIRST survey (off the galactic plane) with the VLBA at 1.4~GHz, reached about 20 per cent \citep{Deller14}.

\begin{figure}
\centering
\includegraphics[width=\linewidth]{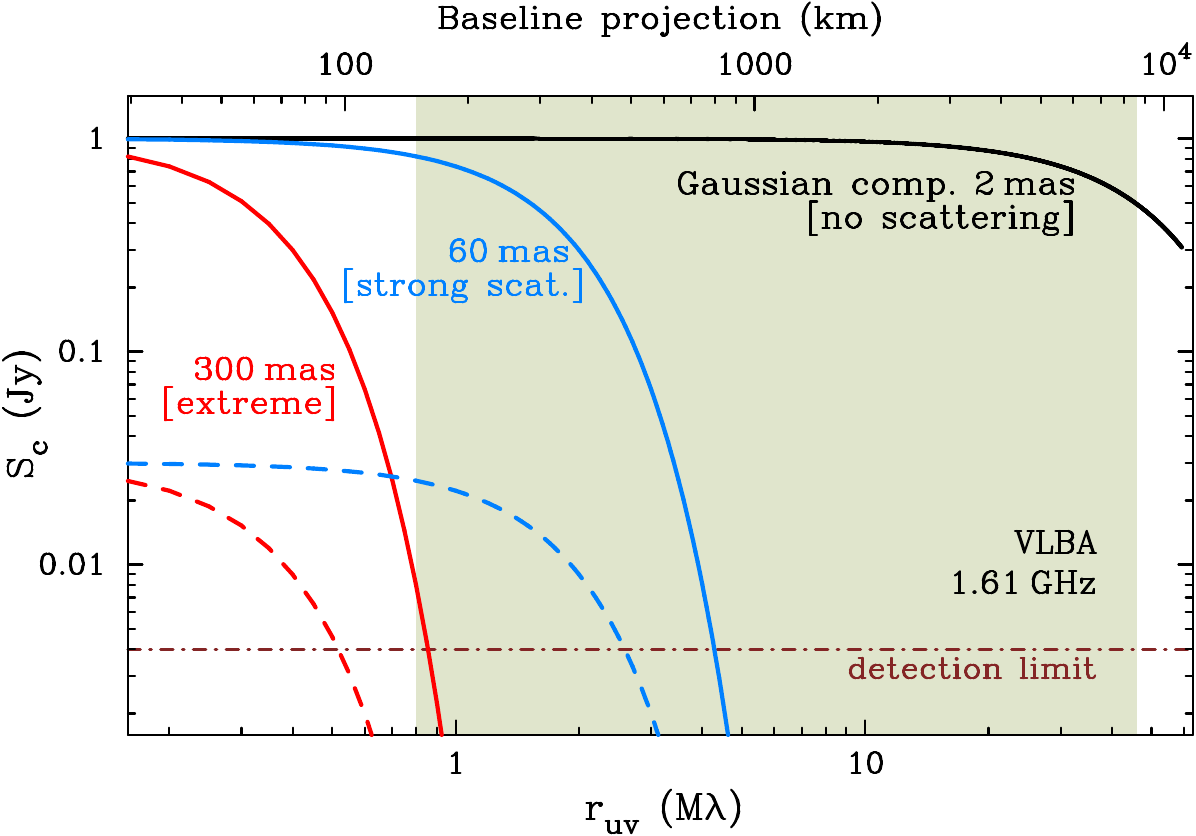}
\caption{Dependence of the correlated flux density on baseline projection for unscattered (black solid line), strongly scattered (blue line), and extremely scattered (red line) Gaussian source with a total flux density 1~Jy observed at L-band. Dashed lines show the case of a weaker source, 30~mJy. Shaded area represents the baseline range provided by the VLBA.}
\label{f:vis_function}
\end{figure}

In \autoref{f:vis_function}, we illustrate the model visibility amplitude of an extended source with a Gaussian brightness distribution, the Fourier transform of which also is a Gaussian \citep{Pearson99}
\begin{gather}
 S_r = S_0\exp\left(\frac{-(\pi\theta r)^2}{4\ln2}\right)\,,
\end{gather}

\noindent where $r$ is the $(u,v)$-radius, $\theta$ is the source FWHM size and $S_r$ is the correlated flux density at a projected spacing $r$. For a bright (e.g. 1~Jy) unscattered source with a typical intrinsic size of about 2~mas at L-band \citep[e.g.,][]{Koryukova22}, the visibility amplitude decreases slowly, ensuring detection even on the longest baselines (black line). The shaded area represents the range of baselines formed with the VLBA. This situation changes dramatically for a scattered source. Strong scattering (blue line) broadens the observed size, causing a rapid drop in correlated flux that falls to the thermal noise level and substantially narrows the $(u,v)$-range, where the source can be detected. The effect is even more severe under extreme scattering (red line). Furthermore, a lower flux density (dashed lines) progressively reduces the likelihood of detection. Consequently, detecting a heavily scattered source, particularly a weak one, is highly challenging. The vast majority of our target sources lie within the regime of strong to extreme scattering for weak sources (see Section~\ref{s:scat_across_GP} and \autoref{f:scat_size_vs_b}). For the successful detection of such sources, which are often weak in flux but strongly scattered, good coverage of short baselines is critical. This can be provided by an interferometer such as e-MERLIN, with the longest baseline being 217~km.

\subsection{Scattering in the quasar J1853$-$0010}
\label{sec:1850-002}

\begin{figure}
\centering
\includegraphics[width=\linewidth]{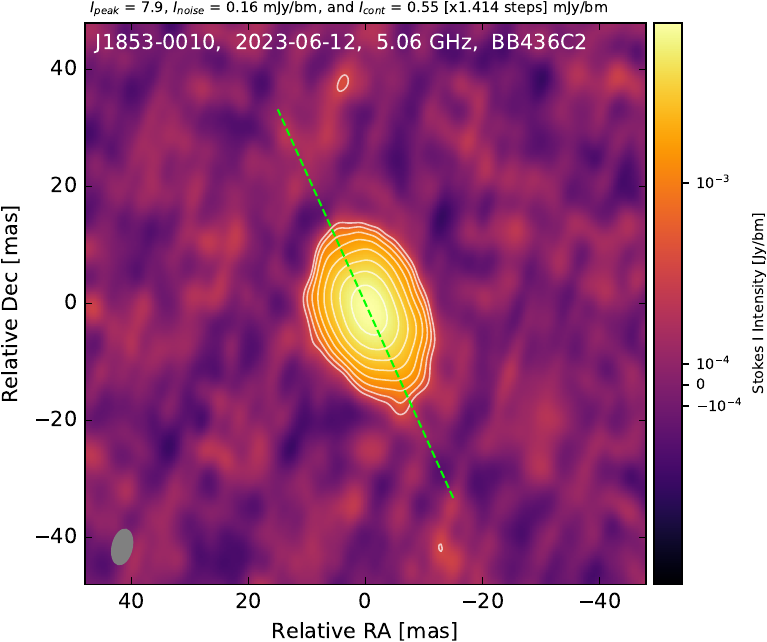}
\caption{Naturally weighted VLBA map of J1853$-$0010 in total intensity at 5.06~GHz (IF4, C-band), with a pixel size of 0.3~mas and Gaussian tapering (value 0.3 at $uv$-radius 50~M$\lambda$) applied. The contours are plotted at increasing powers of $\sqrt{2}$, starting from the corresponding 3.5~rms level. The full-width at half maximum (FWHM) of the restoring beam is shown as a shaded ellipse in the lower left corner. The map parameters are given on the top of the image. The dashed green line denotes the line of constant Galactic latitude ($b=0\fdg468$) at $\mathrm{PA}=27\fdg1$ showing perfect alignment with the direction of the stretched source brightness distribution induced by anisotropic scattering.}
\label{f:P5-1}
\end{figure}

Here we investigate the second brightest object of the field detected at 4.8~GHz, P5-1 ($l=32\fdg969$, $b=-0\fdg468$), the known VLBI source J1853$-$0010 \citep{RFC}. It is about an order of magnitude weaker ($S=52$~mJy of the correlated flux density at the shortest baseline) compared to the strongest source, the quasar 1849+005 (\autoref{fa:radplots_c}) but still bright enough to analyze its size-frequency dependence within C-band, which comprises four adjacent IFs of 512~MHz in total, about 10 per cent of the bandwidth. In \autoref{f:P5-1}, we show the source image at IF4 produced by fitting a single elliptical Gaussian component to the visibility data and performing phase-only self-calibration.

To estimate the size-frequency dependence, source elongation and its direction within the C-band we fitted an elliptical Gaussian component to the observed visibility data. We assumed that within the given band the flux density and the size of the component are power-laws in frequency ($S\sim\nu^{\alpha}$, $\theta \sim \nu^{-k}$) and the source is intrinsically circular unpolarized. We fitted a-priori calibrated but not self-calibrated visibility amplitudes introducing antenna gain amplitudes in the model that are independent among different scans, IFs and polarizations (R/L). Using visibility amplitudes instead of closure amplitudes allows to employ the exact likelihood \citep[which is a \textit{Rice} distribution,][]{doi:https://doi.org/10.1002/9783527617845.ch6} that is important for accurate inference, especially in the case of low S/N data \citep{2022MNRAS.509.3643L}. This also helps to incorporate the prior information on the gain amplitudes \citep{2020ApJ...894...31B}. We considered only visibility data up to some distance in the $uv$-plane, $r_{uv,{\rm max}} \approx 12.5$ M$\lambda$, as due to the large source size, the measurements on longer baselines are uninformative upper limits (Appendix~\ref{sec:visibility_plots}). We also introduced additional per-IF dispersion (``jitter'') to the model to account for possible non-closing errors. To handle the large parameter space and to obtain the uncertainty estimates we followed the Bayesian approach \citep{2008ConPh..49...71T} using \textit{Diffusive Nested Sampling} algorithm \citep{brewer2011} implemented in the \texttt{DNest4} package \citep{JSSv086i07} for sampling the posterior distribution of the model parameters. The summary of the model parameters and corresponding prior distributions are listed in \autoref{tab:intraband_model_priors}.

\begin{table}
\centering
\caption{Model parameters and corresponding prior distributions for the intra-band fitting}
\begin{tabular}{lrl}
\hline\hline\noalign{\smallskip}
Parameter(s) & Prior distribution & Parameter(s) description \\
\hline
$\theta_{\rm maj}(\nu_{\rm max})$ & $\mathcal{LN}(2, 0.75)$ & Major axis FWHM (mas) at $\nu_{\rm max}$ \\
$k$ & $\mathcal{N}(-1, 1)$ & $\theta_{\rm maj}(\nu) \sim \nu^{-k}$ \\
$S(\nu_{\rm max})$ & $\mathcal{LN}(-2, 0.5)$ & Flux density (Jy) at $\nu_{\rm max}$ \\
$\alpha$ & $\mathcal{N}(0, 0.5)$ & $S(\nu) \sim \nu^\alpha$ \\
$e$ & $\mathcal{U}(0, 1)$ & Ellipticity - $\theta_{\rm min}/\theta_{\rm maj}$ \\
PA & $\mathcal{U}(0, \upi)$ & Major axis position angle \\
$|g^{\rm L/R}_{t,s,i}|$ & $\mathcal{N}(1, 0.1)$ & Gain amp. of ant. $t$ for scan $s$, IF $i$ \\
$\sigma_{{\rm jitter},i}$ & $\mathcal{LN}(-5, 1)$ & Additional dispertion (Jy) for IF $i$ \\
\hline
\end{tabular}
\begin{tablenotes}
\item
$\mathcal{N}$, $\mathcal{LN}$ and $\mathcal{U}$ are Normal, Log-Normal and Uniform distributions, while $\nu_{\rm max}$ corresponds to the highest frequency IF in a given band.\\
\end{tablenotes}
\label{tab:intraband_model_priors}
\end{table}

\begin{figure}
\centering
\includegraphics[width=0.82\columnwidth]{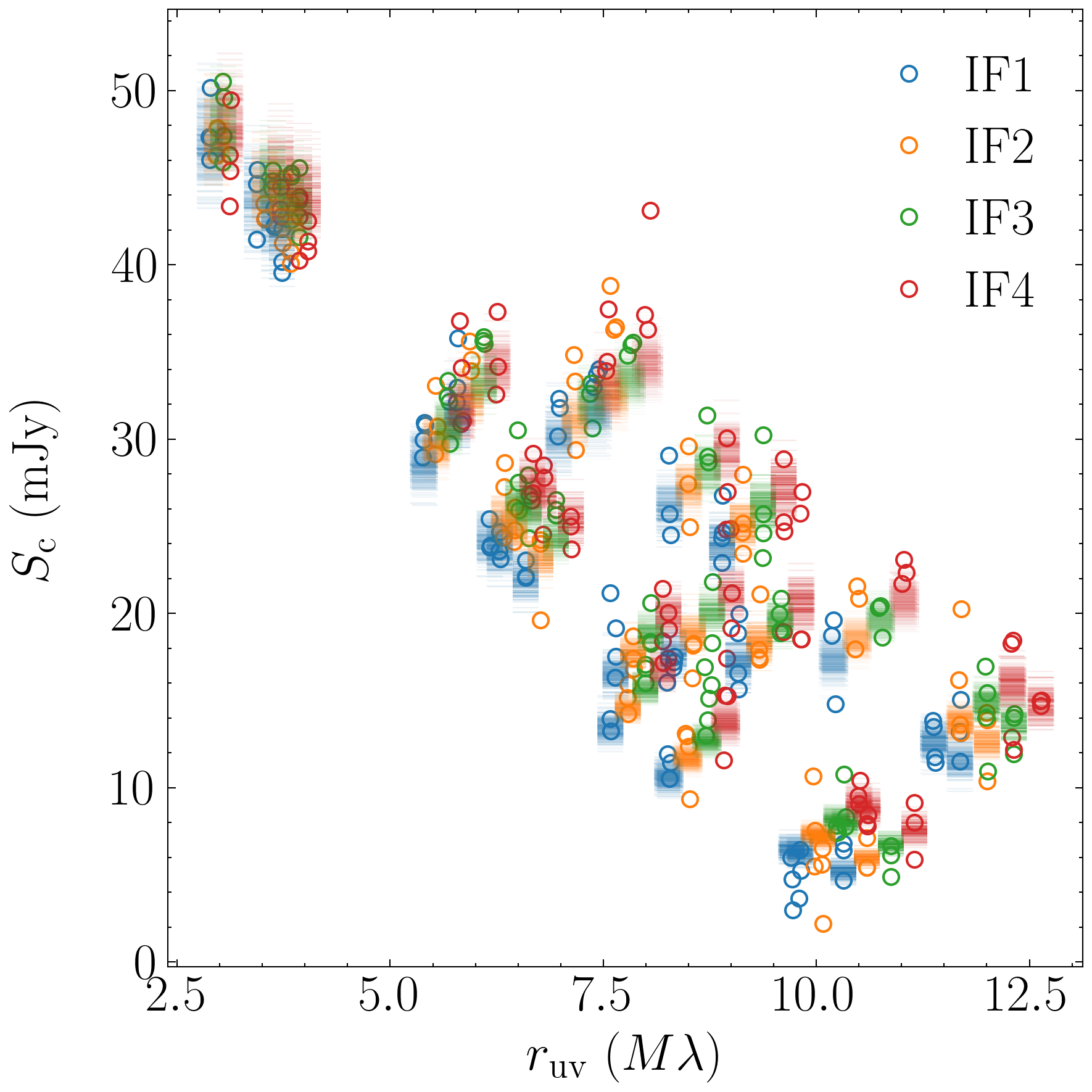}
\caption{C-band observed a-priori calibrated but not self-calibrated visibility amplitude (circles) for Stokes I vs. distance in $uv$-plane for quasar 1853$-$0010 (Section~\ref{sec:1850-002}). Horizontal line segments show predictions for 200 samples of the posterior distribution of the scattered elliptical Gaussian model parameters without antenna gains contribution. Different colours show different IFs. Characteristic amplitude shift of the different IFs data is evident, which is a result of the scattered size frequency dependence.}
\label{f:1850-002_Cband_Ifit}
\end{figure}

\begin{figure*}
\centering
\includegraphics[width=0.7\linewidth]{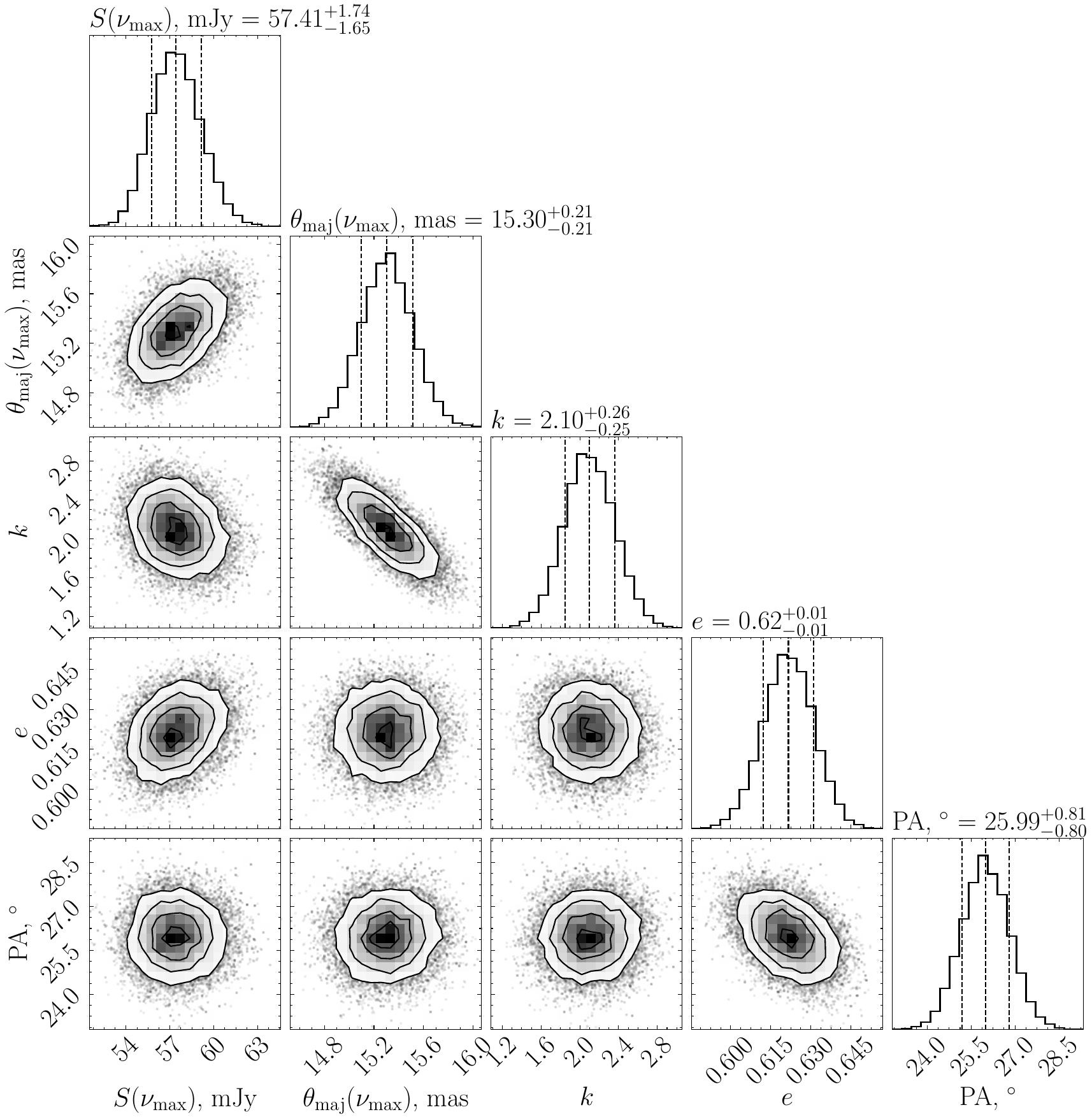}
\caption{Posterior distribution of the scattered elliptical Gaussian component parameters for quasar J1850$-$002 obtained from C-band data (Section~\ref{sec:1850-002} and Table~\ref{tab:intraband_model_priors}). Two and one dimensional projections (i.e.\ marginalized posteriors) of the distribution are shown. Vertical dashed lines in histograms and values above show 15, 50 and 84 per cent quantiles of the distribution.}
\label{f:1850-002_corner}
\end{figure*}

The observed a-priori calibrated but not self-calibrated data of quasar J1853$-$0010 and Gaussian model predictions for the C-band are shown in \autoref{f:1850-002_Cband_Ifit}. Here we plot the model prediction without contribution of the gain amplitudes to display its characteristic scattering-induced frequency dependence across IFs. The estimated unaccounted dispersion (``jitter'') is $\approx1$ mJy for each IF that is less than the median thermal uncertainty of the data points ($\approx 2$ mJy), i.e. the model describes the observed data well. The posterior distribution of the source model parameters is presented in \autoref{f:1850-002_corner}. The spectral index $\alpha$ is only weakly constrained by such narrow frequency band data so we did not include it in the plot. The source is significantly ($e=0.62\pm0.01$) elongated almost exactly along the Galactic plane (${\rm PA}=26\fdg0\pm0\fdg8$ vs ${\rm PA}_{b={\rm const}}=27\fdg1$) as shown in \autoref{f:P5-1}. The frequency dependence of the source size with the index $k=2.10\pm0.26$ strongly favors scattering. More precise estimates of $k$ were obtained for the brighter quasar 1849$+$005, for which several data sets covering a wide frequency range were analyzed (Pushkarev et al., in prep.). The observed source morphology stretched along the line of constant Galactic latitude indicates that (i) the source emission is subject to anisotropic scattering, (ii) the scattering screen(s) is oriented orthogonal to the Galactic plane; thus, it can be radio filaments related to pulsars \citep[e.g,][]{Churazov24} or flows of hot ionized gas moving upward from the Galactic disk in regions with active star formation \citep{Churazov24_North_Polar_Shpur}. Similar observational properties were previously found in the quasars 2023+335 \citep{Pushkarev13} and 2005+403 \citep{Koryukova22}.

\subsection{Scattering power across the Galactic plane}
\label{s:scat_across_GP}

Early VLBI observations of eight sources in the Cygnus region ($l\approx 75^\circ\pm8^\circ$) demonstrated that four of them are clearly subject to scattering, and the stripe of angular broadening is confined within a few degrees from the Galactic plane \citep{Fey89}. More recent analysis of VLBI core sizes derived for thousands of AGNs that cover almost the entire sky, except the Southern Celestial Pole region, showed that scattering manifests itself statistically within $|b|\leq10^\circ$ \citep{Pushkarev15,Koryukova22}, with significant enhancement toward the Galactic center region including the bar.

\begin{figure*}
\centering
\includegraphics[width=0.8\linewidth]{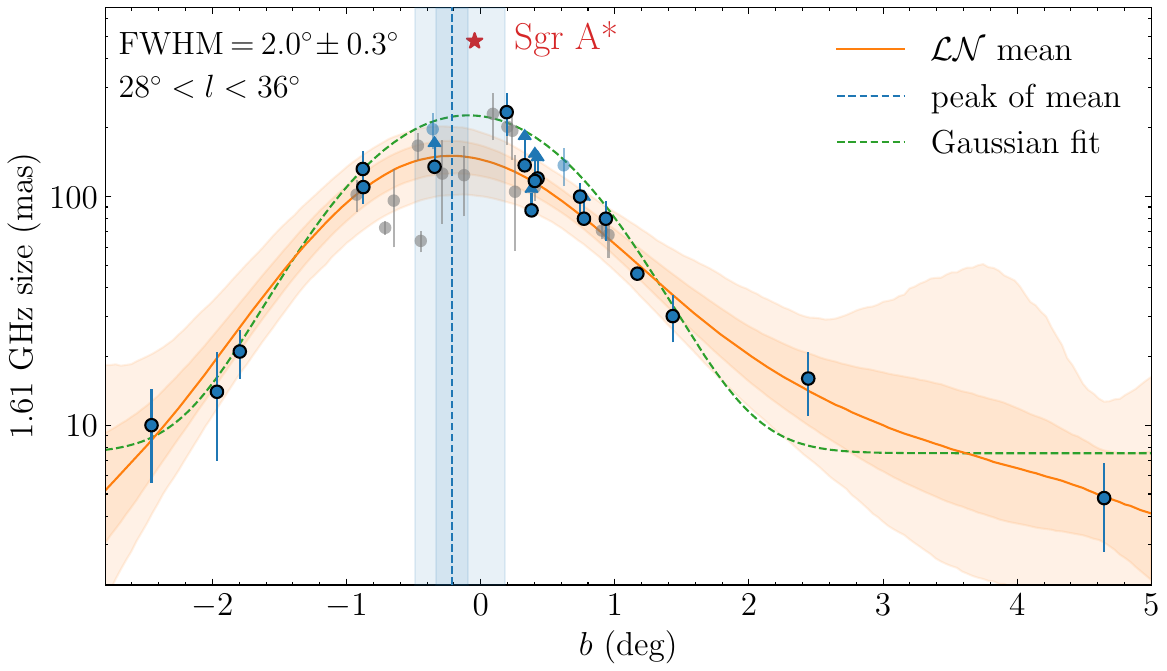}
\caption{Observed size along the major axis of the fitted elliptical Gaussian component for sources detected at L-band (1.61~GHz; blue circles) and S-band (2.3~GHz; light blue circles; re-scaled to 1.61~GHz assuming $\theta\sim\lambda^2$ dependence) as a function of their Galactic latitude $b$ fitted with a Log-Normal Gaussian Process mean model (Section~\ref{s:scat_across_GP}). The points with arrows are lower limits, with the arrow length corresponding to estimated error. The orange curve and bands are the expected mean value of the size at given $b$ and its 64 and 95 per cent credibility intervals (CIs). The blue vertical dashed line and vertical bands are the position of the maximum of the mean and its 64 and 95 per cent CIs. The dashed green line is the Gaussian with amplitude $216\pm30$~mas, location $-0\fdg09\pm0\fdg06$, FWHM $1\fdg73\pm0\fdg14$ and offset $7.4\pm2.4$ mas, obtained by fitting a Gaussian to the 64-th percentile of the Log-Normal distribution that describes the data at each $b$ (i.e. consistent with model from Section~\ref{s:scat_across_GP}). The plot is complemented by angular sizes derived from C-band (4.87~GHz) and re-scaled to L-band (gray points). Red star marks Sgr~A* position \citep{Sgr_A_position} and its angular size in mas along the major axis based on the relation $(1.380\pm0.013)\lambda^2_\text{cm}$ derived by \citet{Johnson18}. The right most point with $b=4\fdg65$ marks the phase calibrator J1833+0015.}
\label{f:scat_size_vs_b}
\end{figure*}

Out of 32 sources detected by our VLBA observations at L, S, and C-bands, we were able to estimate the observed angular size $\theta$ for 27 of them by fitting an elliptical Gaussian component in the Caltech {\sc difmap} package \citep{DIFMAP} to the initially calibrated, but not self-calibrated data. The remaining five sources, all at L-band, are either substantially weaker and/or extremely strongly scattered, making a reliable assessment of their sizes challenging. In the latter case, baselines shorter than available are needed to trace the visibility function by detecting the corresponding correlated flux densities to properly measure the source size. There were seven more quasars, including the phase calibrator J1833$+$0115, located outside but near the pilot field, at $|b|>1^\circ$. Each of them, except J1833$+$0115, had only a single scan, but since they are relatively bright, with a correlated flux density at the shortest baselines $>0.1$~Jy, their angular sizes were also reliably measured. Thus, in total, we have 34 sources with derived $\theta$. 

\begin{figure}
\centering
\includegraphics[width=\linewidth]{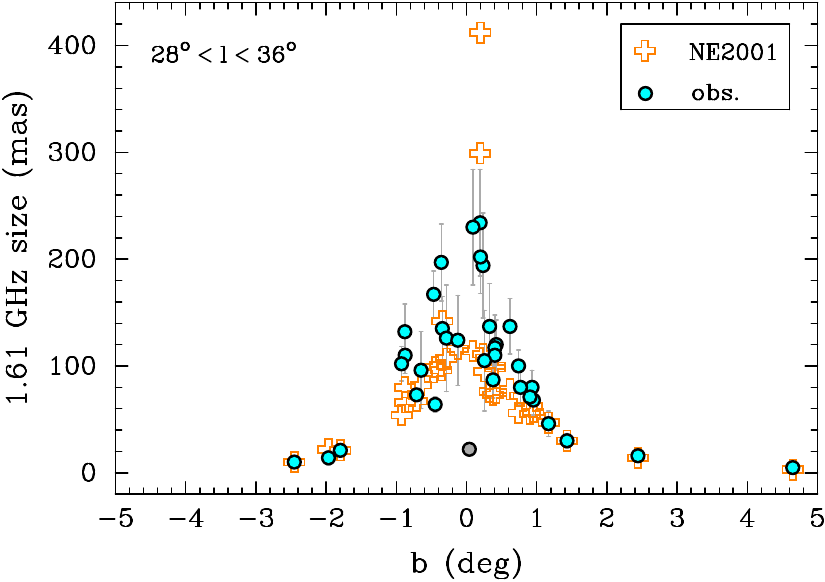}
\caption{Measured sizes as in \autoref{f:scat_size_vs_b} in comparison with the predictions of the NE2001 model model scaled to the same frequency of 1.61~GHz. While there is excellent agreement for sizes up to 50~mas for the sources with $|b|>1^\circ$, there are sources with the measured size significantly larger than the expected size for expected sizes between 50 and 150~mas. Grey-filled circle marks the pulsar B1845$-$01, see Section~\ref{s:G_detection}.}
\label{f:obs-vs-model}
\end{figure}

\begin{figure*}
\centering
\includegraphics[width=2\columnwidth]{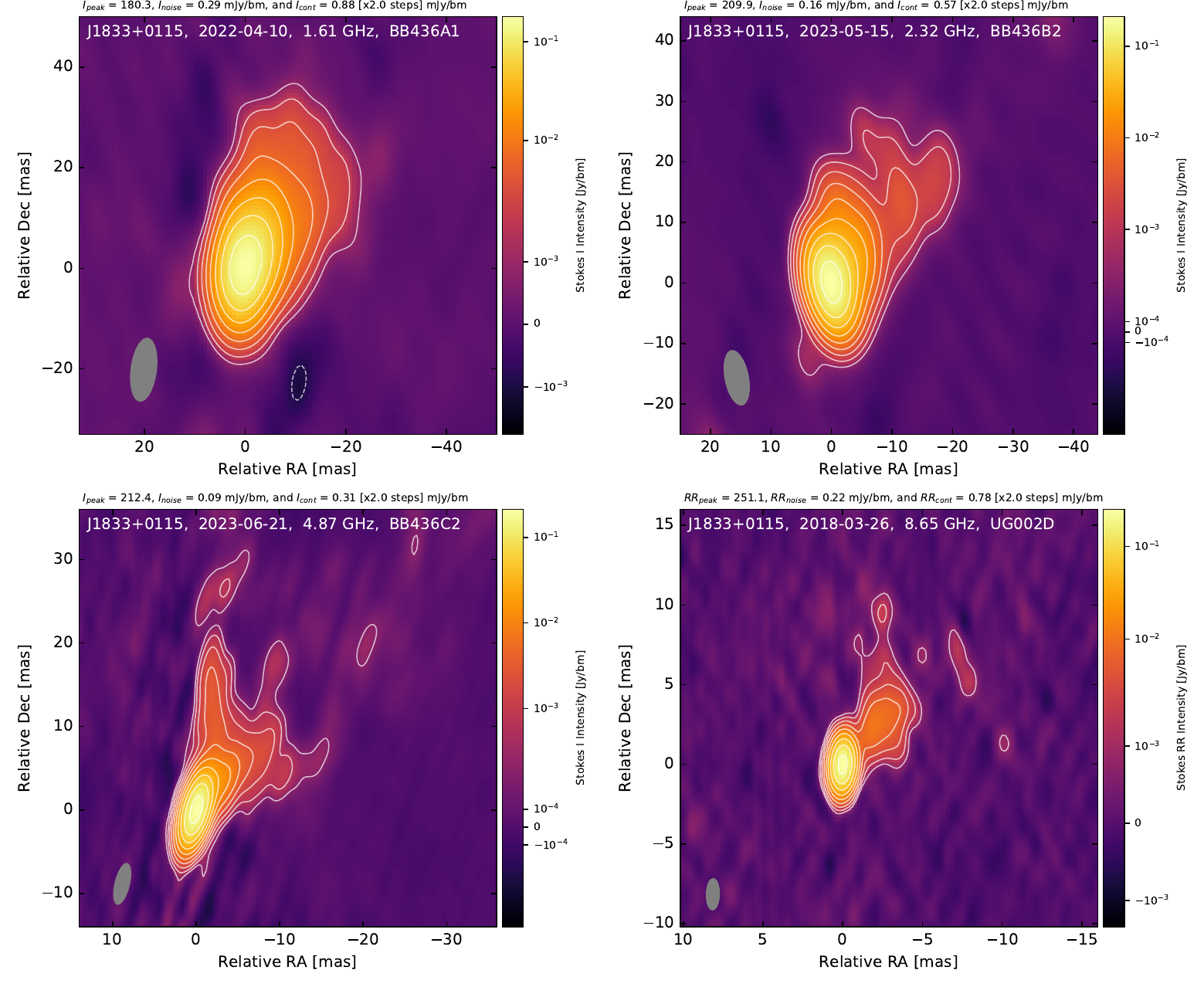}
\caption{Naturally weighted VLBA maps of J1833+0115 in total intensity at 1.61~GHz (upper left), 2.32~GHz (upper right), 4.87~GHz (lower left) and in Stokes RR at 8.65~GHz (lower right), with a cell size of 0.8, 0.5, 0.3 and 0.2~mas per pixel, respectively. The contours are plotted at increasing powers of 2, starting from the corresponding 3.5~rms levels. The full-width at half maximum (FWHM) of the restoring beam is shown as a shaded ellipse in the lower left corner. The map parameters are given on the top of each image.}
\label{f:phase_cal}
\end{figure*}

In \autoref{f:scat_size_vs_b}, we show the angular sizes measured at or re-scaled from S or C to L-band (1.61~GHz) as a function of the Galactic latitude. We fitted the dependence of the observed angular sizes on $b$ for 20 sources distributed nearly uniformly across the Galactic plane and covering a range of Galactic latitudes from $-2\fdg5$ to $4\fdg6$ (\autoref{f:scat_size_vs_b}). We assumed that at each $b$ the distribution of the observed sizes is \textit{LogNormal} \citep{aitchison1957lognormal}: $\theta(b) \sim \mathcal{LN}(\mu_{\theta}(b), \sigma_{\theta})$, where the location parameter $\mu_{\theta}$ depends on the $b$ via a Gaussian Process (GP): $\mu_{\theta}(b) \sim \mathcal{GP}(\mu_{\rm G}, C_{\rm G})$ with the GP mean function $\mu_{\rm G}$ fixed to the logarithm of the typical size at L-band and squared exponential GP kernel $C_{\rm G}$ \citep{RasmussenW06}. The FWHM of the fitted dependence is $2\fdg0\pm0\fdg3$, consistent with a typical width of roughly $1^\circ-3^\circ$ of the thin disk of the Milky Way \citep{Cordes84,NE2001,Yao17}, where most of the turbulent medium resides. Surprisingly, it is still narrow, despite being directionally close to the Galactic bar. The fitted curve peaks at $-0\fdg21$, which is consistent with 0 at the level $1.4\sigma$. However, it is interesting to note that this slight difference from zero would be expected, since the Sun is located slightly above the Galactic mid-plane. The estimates range from 5.5~pc \citep{Reid2019} to 20.8~pc \citep{Bennett2019}. For a height of 20.8 pc, a line-of-sight at $b=\-0\fdg21$ toward the inner Galaxy would cross the Galactic mid-plane at a distance of roughly 5.5~kpc. This corresponds roughly to the location of the end of the bar, where a large fraction of the scattering is expected to occur. The non-zero width $\sigma_{\theta} = 0.27\pm0.14$
of the $\log\theta(b)$ distribution on a given $b$ reflects that the turbulent ISM is clumpy and fractal in nature on the corresponding scales \citep{Combes00}, and the scattering power can vary substantially. Alternatively, lower values $\theta$ may represent lower limits, as the shortest baseline projections are limited by about 180~km between the Los Alamos and Pie Town (PT) VLBA stations. Inclusion of a single-dish VLA antenna (Y1) would greatly help, providing sufficiently shorter ($\sim$20~km) baseline projections to PT. The observed increase of $\theta$ toward $b\approx0^\circ$ is mainly driven by longer path lengths through the most dense and turbulent ionized interstellar medium in the Galaxy, indicating that the lines of sight at $|b|<1^\circ$ likely cross multiple scattering screens. In \autoref{fa:vis_func_L_all_sources}, we present model visibility functions of all target sources at L-band assuming that their angular sizes follow the derived distribution $\theta(b)$ and using the total flux densities from the GLOSTAR catalog \citep{Dzib23}. This plot shows that most of the sources, being weak but strongly scattered, are below the detection limit when we lack baselines shorter than those offered by the VLBA, approximately $150$~km.

We compared the source sizes $\theta$ at 1.61~GHz derived from our observations with those of the NE2001 model \citep{NE2001} re-scaled to the same frequency (\autoref{f:obs-vs-model}). For all sources with $|b|>1^\circ$, the agreement is quite impressive, while for sources at $|b|<1^\circ$ the observed sizes are on average larger than the model ones, except two sources at the smallest deviations from the Galactic plane.

\subsection{Scattering in phase calibrator J1833+0115}

The bright source J1833+0115 ($l=31\fdg965$, $b=4\fdg648$), acting as a phase calibrator for our observations, is a few degrees apart from the pilot sky region. The source was observed every other scan over 6~h, with a total tracking time of $\sim$2~h at each frequency band. In \autoref{f:phase_cal}, we present restored CLEAN maps at 1.61, 2.32, 4.87~GHz in total intensity based on our observations and at 8.65~GHz in RR corelation from geodetic VLBA observations\footnote{\url{ http://astrogeo.org/vlbi_images/}}. CLEANing, phase and amplitude self-calibration were performed in {\sc difmap}. The source manifests a typical parsec-scale morphology of AGN, with a bright core and one-sided jet developing in a north-west direction. Beyond 5~mas from the core the jet becomes much wider and edge-brightened, as revealed at 4.87~GHz (\autoref{f:phase_cal}, bottom left), with an apparent opening angle of $\sim$$45^\circ$. At lower frequencies, especially at 1.61~GHz the jet is traced up to $\sim$40~mas from the core and its cross-section is filled and is no longer limb-brightened.

\begin{figure}
\centering
\includegraphics[width=\linewidth]{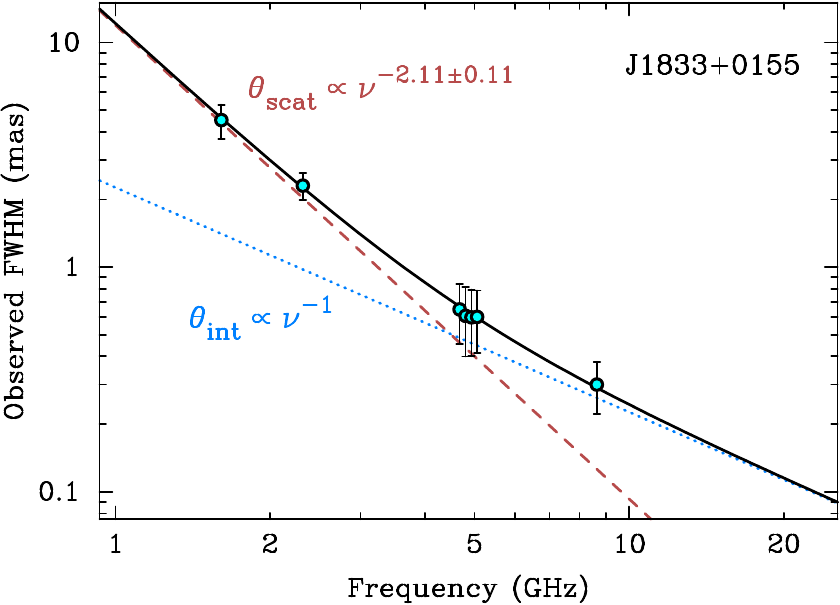}
\caption{Observed angular size of the VLBI core component of J1833+0115 versus frequency. The solid line shows the fit of Eq.~(\ref{eq:observed_size}), while the dashed and dotted lines represent the inferred intrinsic and scattered size, respectively.}
\label{f:theta_vs_nu}
\end{figure}

In \autoref{f:scat_size_vs_b}, the phase calibrator J1833+0115, which is at the largest separation from the Galactic plane, shows the smallest angular size. However, the source is still subject to scattering, as is evident from \autoref{f:theta_vs_nu}, where we plot the angular size of the core component versus the observing frequency and fit the following dependence
\begin{gather}
    \theta_\mathrm{obs}^2 = \left(\theta_\mathrm{int}\,\nu^{-1}\right)^2 + \left(\theta_\mathrm{scat}\,\nu^{-k}\right)^2 \,,
	\label{eq:observed_size}
\end{gather}
where $\theta_\mathrm{int}$ and $\theta_\mathrm{scat}$ are the intrinsic and scattered angular sizes of a source at 1~GHz, respectively, and $k$ is the scattering index. We assume that $\theta_\mathrm{int}\propto\nu^{-1}$, i.e., the jet shape is conical on the scales where the core feature is located at different frequencies \citep{BK79}. The best-fit values derived are $\theta_\mathrm{int}=2.26\pm0.18$~mas, $\theta_\mathrm{scat}=11.95\pm0.79$~mas, and $k=2.11\pm0.11$. The latter implies that the source emission is subject to scattering. As apparent from \autoref{f:theta_vs_nu}, scattering dominates in angular broadening at frequencies below $\sim$4~GHz. The value of the $k$-index lying between 2.0 and 2.2 indicates that the scattering is in the regime when the inner scale of turbulence $r_\mathrm{in}$ is comparable to the diffractive scale $r_\mathrm{diff}$, assuming a Kolmogorov spectrum of electron-density inhomogeneities \citep{Wilkinson94}. We can estimate $r_\mathrm{diff}$, since it is a baseline at which visibility decreases by a factor of $1/\sqrt{e}$, which yields about 8~M$\lambda$ at L-band (\autoref{fa:radplot-l-phas-cal}). Thus, we obtain $r_\mathrm{in}\approx1500$~km as an estimate on a dissipation scale of turbulence. We also note that the inferred $\theta_\mathrm{int}$ agrees well with a typical source size at high Galactic latitudes where scattering is virtually not present, e.g. $\sim$1~mas at 2.3~GHz \citep{Pushkarev15,Koryukova22}.

\subsection{Detecting a Galactic source}
\label{s:G_detection}

\begin{figure}
\centering
\includegraphics[width=\linewidth]{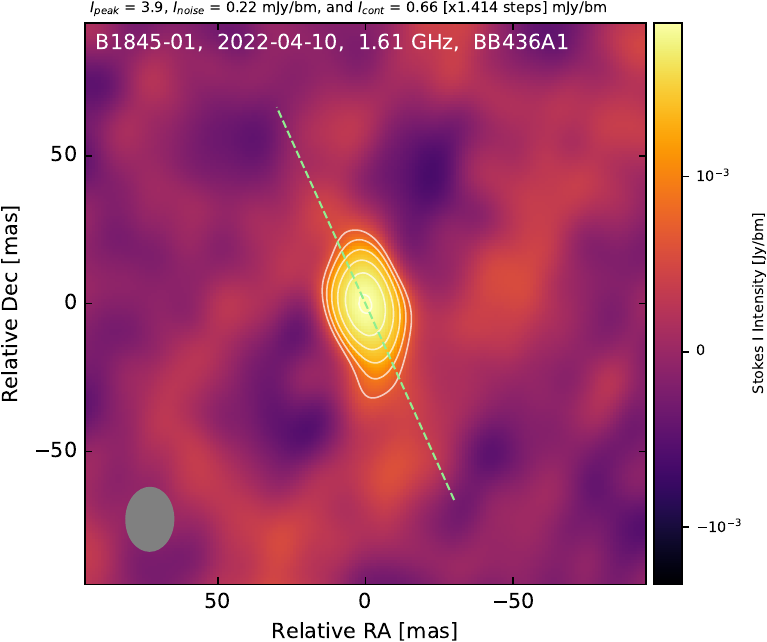}
\caption{Naturally weighted VLBA map of a Galactic source, the pulsar B1845$-$01 (P47-40) in total intensity at 1.61~GHz, with a pixel size of 1~mas and tapering applied. The contours are plotted at increasing powers of $\sqrt{2}$, starting from the corresponding 3~rms level. The full-width at half maximum (FWHM) of the restoring beam is shown as a shaded ellipse in the lower left corner. The map parameters are given on the top of the image. The dashed green line denotes the line of constant Galactic latitude ($b=0\fdg039$) at $\mathrm{PA}=27\fdg1$ showing good alignment with the direction of the stretched source brightness distribution induced by anisotropic scattering.}
\label{f:P47-40}
\end{figure}

Our target list comprising 1210 sources includes three pulsars. Among those 18 objects detected at L-band, there is a source P47-40 located at $l=31\fdg339$, $b=0\fdg039$ (\autoref{f:sky_coverage}, upper panel, \autoref{fa:radplots_l}), which is the pulsar B1845$-$01, at a distance of about 5~kpc \citep{lwy16,Yao17}. In \autoref{f:P47-40}, we show its restored image, with a peak-to-noise ratio of 23. The revealed source brightness distribution is elongated, reflecting the scattering kernel since the background source is effectively point-like. The position angle of the apparent brightness anisotropy is closely aligned with the direction of the $b=\textrm{const}$ line. We note that the total flux (4.4~mJy) of the source and its angular size (22~mas) derived by fitting an elliptical Gaussian component could be lower limits as their derived values are affected by a lower limit on baseline projection, 0.84~M$\lambda$. However, detection the source at such a small Galactic latitude and its relatively small angular size (roughly an order of magnitude smaller than expected from the fit in \autoref{f:scat_size_vs_b})  indicates that the majority of the heavy scattering is occurring behind the pulsar.

\section{Conclusions}  
\label{sec:conclusion}

Utilizing multi-frequency (1.6, 2.3, and 4.8~GHz) VLBA observations, we have studied scattering properties in the 16 square degrees pilot region of the GLOSTAR survey toward the end of the Galactic bar that covers 1210 target sources within the stripe of $28^\circ<l<36^\circ$ and $|b|<1^\circ$. The main results of our project are as follows.
\begin{enumerate}
\item[1.] The source emission is subject to heavy scattering as indicated by (i) their low detection rate, 1.5, 3.4 and 9.2 per cent at 1.6, 2.3 and 4.8~GHz, respectively, which increases with observing frequency, nearly scaling as $\nu^2$ dependence; (ii) a statistically significant tendency for sources to be detected at higher absolute values of Galactic latitudes; (iii) the detected sources avoid sky areas covered by the known highly-turbulent supernova remnants, where the power of scattering is expected to be even higher.
\item[2.] We detected anisotropic scattering in the extragalactic radio source 1850$-$002 (J1853$-$0010). Its brightness distribution at C-band induced by refractive-dominated propagation effect is stretched ($\text{PA}=26\fdg0\pm0\fdg8$) along the line of constant Galactic latitude ($\text{PA}=27\fdg1$). The corresponding power-index of the scattered size inferred within the C-band is $2.10\pm0.26$.
\item[3.] The VLBI core component of the phase calibrator, quasar J1833+0115, located at $b=4\fdg65$ is found to be isotropically scattered. The scattering index derived from multi-frequency VLBA observations at 1.6, 2.3, 4.8, and 8.6~GHz is $k=2.11\pm0.11$, and the dependence between the observed angular size against frequency shows that scattering dominates at $\nu\lesssim4$~GHz. The inferred $k$-value suggests that the inner scale of turbulence is of the order of $10^3$~km, assuming a Kolmogorov turbulence spectrum.
\item[4.] The scattering power across the Galactic plane based on the derived source angular sizes is not uniform and can be approximated by a Gaussian with a width of about $2^\circ$ that peaks at the Galactic mid-plane, indicating that lines of sight within $|b|<1^\circ$ cross multiple regions with turbulent plasma leading to a cumulative angular broadening. The observed sources sizes are well consistent with those from the NE2001 model for objects with $|b|>1^\circ$, while at $|b|<1^\circ$ the observed sizes are on average higher than the predicted ones, except for two sources at $b\approx0^\circ$. 
\item[5.] At 1.61~GHz, we detected one Galactic source, the pulsar B1845$-$01. It also shows anisotropic scattering along the $b=\textrm{const}$ line. The angular source size is roughly an order of magnitude smaller than those of other detected objects, strongly suggesting that B1845$-$01 is located closer to us than the majority of the scattering media toward the pilot sky region we probed with our VLBA observations.
\end{enumerate}

To significantly improve the source detection rate given strong scattering, future observations must include shorter baselines. This can be achieved either by adding a single VLA antenna (Y1) to the VLBA, creating projected spacings down to about 20~km with Pie Town, or by utilizing the MERLIN network, which provides baselines ranging from approximately 11 to 217~km.

\section*{Acknowledgements}

We are grateful to the anonymous referee, whose comments have helped us to improve the manuscript.
This research emerged from discussions at the online EVN min-symposium in July 2021. This study was supported by the Russian Science Foundation grant 25-22-00152.
Y.Y.K.\ was supported by the MuSES project, which has received funding from the European Union (ERC grant agreement No 101142396). Views and opinions expressed are however those of the author(s) only and do not necessarily reflect those of the European Union or ERCEA. Neither the European Union nor the granting authority can be held responsible for them.
A.V.P.\ is a postdoctoral fellow at the Black Hole Initiative, which is funded by grants from the John Templeton Foundation (grants 60477, 61479, 62286) and the Gordon and Betty Moore Foundation (grant GBMF-8273).
The views and opinions expressed in this work are those of the authors and do not necessarily reflect the views of these Foundations.
M.M.L. was supported by the FONDECYT iniciaci{\'o}n project 11251078. 
This work made use of the Swinburne University of Technology software correlator \citep{DiFX}, developed as part of the Australian Major National Research Facilities Programme and operated under licence.
S.A.D.\ contributed to the preparation of the observing proposal and to early calibration work carried out during the initial phase of the project, and later performed an independent analysis relevant to this study.
N.R. acknowledges support from the United States-India Educational Foundation through the Fulbright Program.

This research made use of \textit{pymc} probabilistic programming library for Python \citep{10.7717/peerj-cs.1516} and Python packages: \textit{Numpy} \citep{harris2020array}, \textit{Scipy} \citep{2020SciPy-NMeth}. \textit{Pandas} \citep{the_pandas_development_team_2025_15597513}, \textit{Matplotlib} \citep{Hunter:2007}, \textit{SciencePlots} \citep{SciencePlots}, \textit{corner} \citep{corner}.

\section*{Data Availability}

The interferometric data used in this paper are publicly available from the NRAO data archive\footnote{\url{https://data.nrao.edu}}. The fully calibrated data will be shared on reasonable request to the corresponding author.

\bibliographystyle{mnras}
\bibliography{galactic_bar}

\appendix

\section{Visibility functions from observations and simulations}
\label{sec:visibility_plots}

\begin{figure*}
\centering
\includegraphics[width=\linewidth]{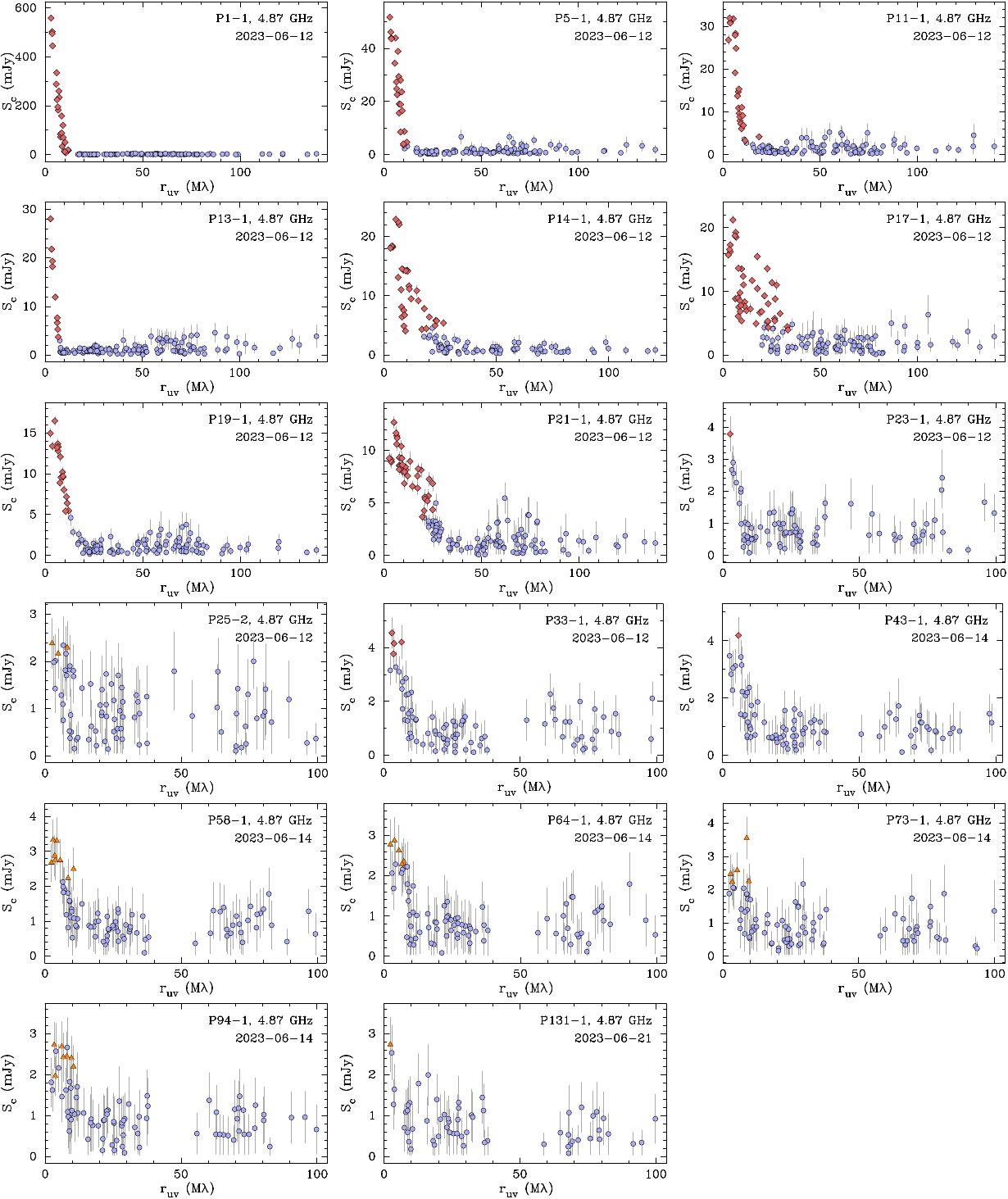}
\caption{Correlated flux density, $S_\text{c}$, as a function of baseline projection $r_{uv}$ for 17 sources detected at C-band. Every point marks a coherent average over a scan length of $\sim$2.5~min and over IFs on an individual interferometer baseline. Red diamonds represent visibilities with $\text{S/N}>6$, orange triangles mark data with $\text{S/N}>4$, lavender points show the measurements with a lower significance. Data were phase-calibrated to a strong nearby calibrator. Error bars represent statistical $1\sigma$ uncertainties. Each plot legend shows the observing epoch, frequency and a source name in the internal notation PXX-X. Rapid decrease of the visibility function amplitudes with baseline length indicates that the source is heavily resolved.}
\label{fa:radplots_c}
\end{figure*}

\begin{figure*}
\centering
\includegraphics[width=\linewidth]{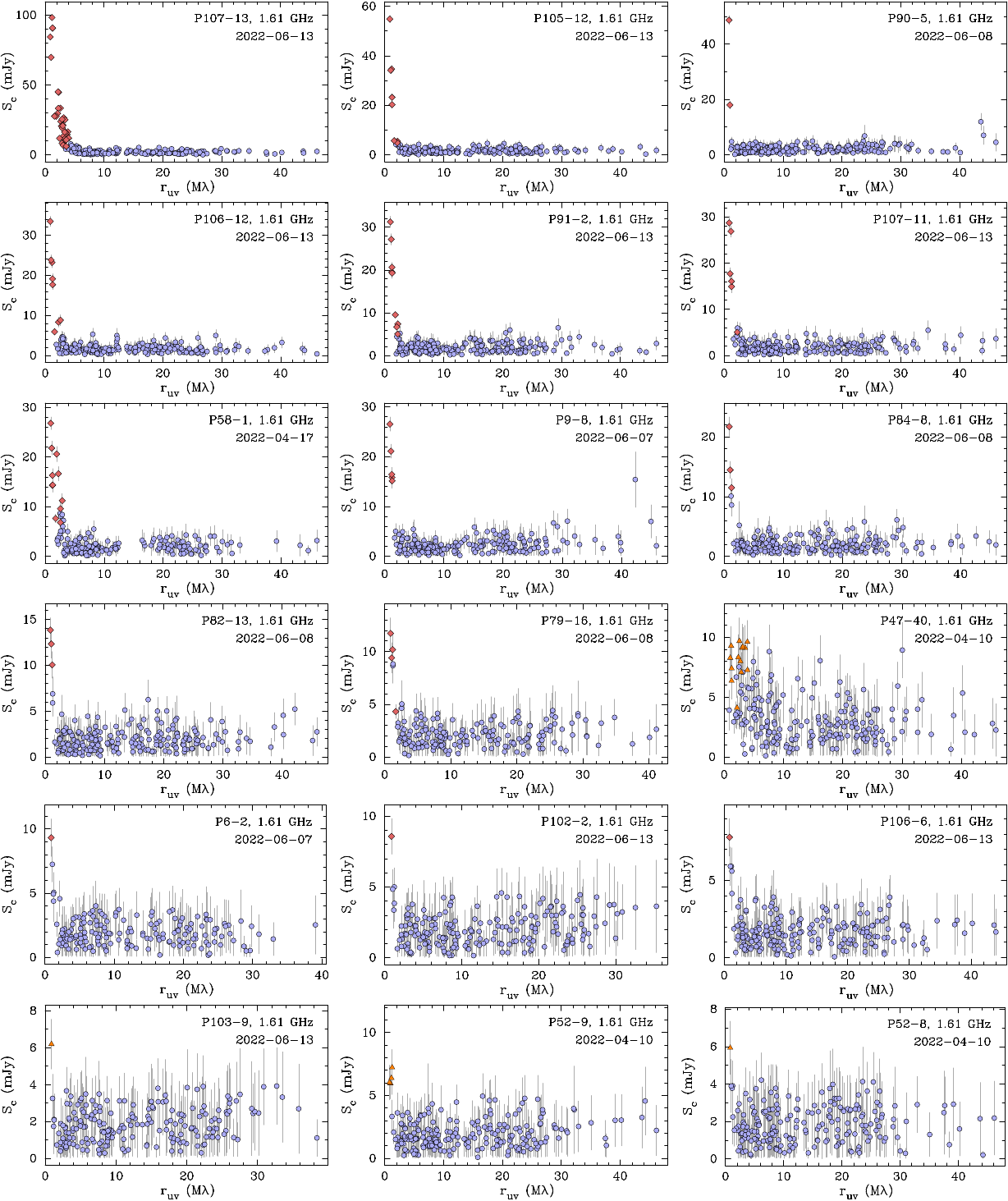}
\caption{Correlated flux density, $S_\text{c}$, as a function of baseline projection $r_{uv}$ for 18 sources detected at L-band. Every point marks a coherent average over a scan length of $\sim$2.5~min and over IFs on an individual interferometer baseline. Red diamonds represent visibilities with $\text{S/N}>6$, orange triangles mark data with $\text{S/N}>4$, lavender points show the measurements with a lower significance. Data were phase-calibrated to a strong nearby calibrator. Error bars represent statistical $1\sigma$ uncertainties. Each plot legend shows the observing epoch, frequency and a source name in the internal notation PXX-X. Rapid decrease of the visibility function amplitudes with baseline length indicates that the source is heavily resolved.}
\label{fa:radplots_l}
\end{figure*}

\begin{figure*}
\centering
\includegraphics[width=\linewidth]{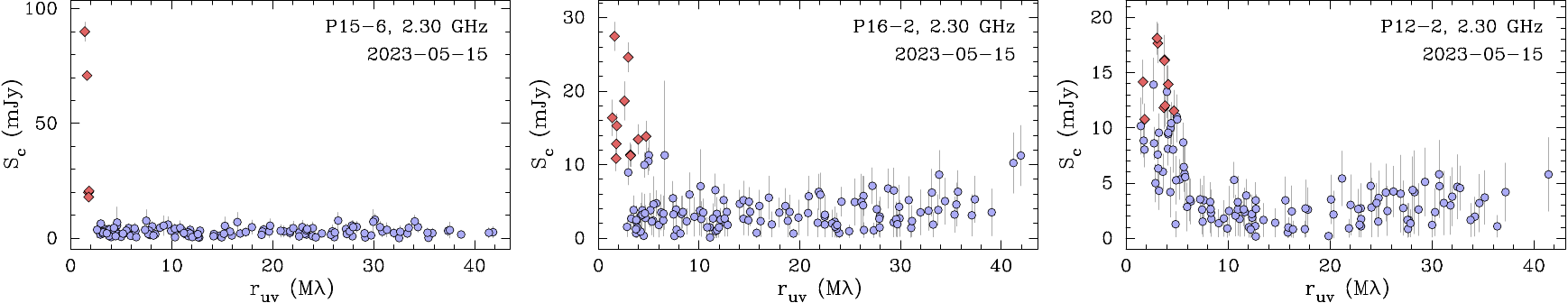}
\caption{Correlated flux density, $S_\text{c}$, as a function of baseline projection $r_{uv}$ for 3 sources detected at S-band. Every point marks a coherent average over a scan length of $\sim$2.5~min and over IFs on an individual interferometer baseline. Red diamonds represent visibilities with $\text{S/N}>6$, orange triangles mark data with $\text{S/N}>4$, lavender points show the measurements with a lower significance. Data were phase-calibrated to a strong nearby calibrator. Error bars represent statistical $1\sigma$ uncertainties. Each plot legend shows the observing epoch, frequency and a source name in the internal notation PXX-X. Rapid decrease of the visibility function amplitudes with baseline length indicates that the source is heavily resolved.}
\label{fa:radplots_s}
\end{figure*}

\begin{figure*}
\centering
\includegraphics[width=0.7\linewidth]{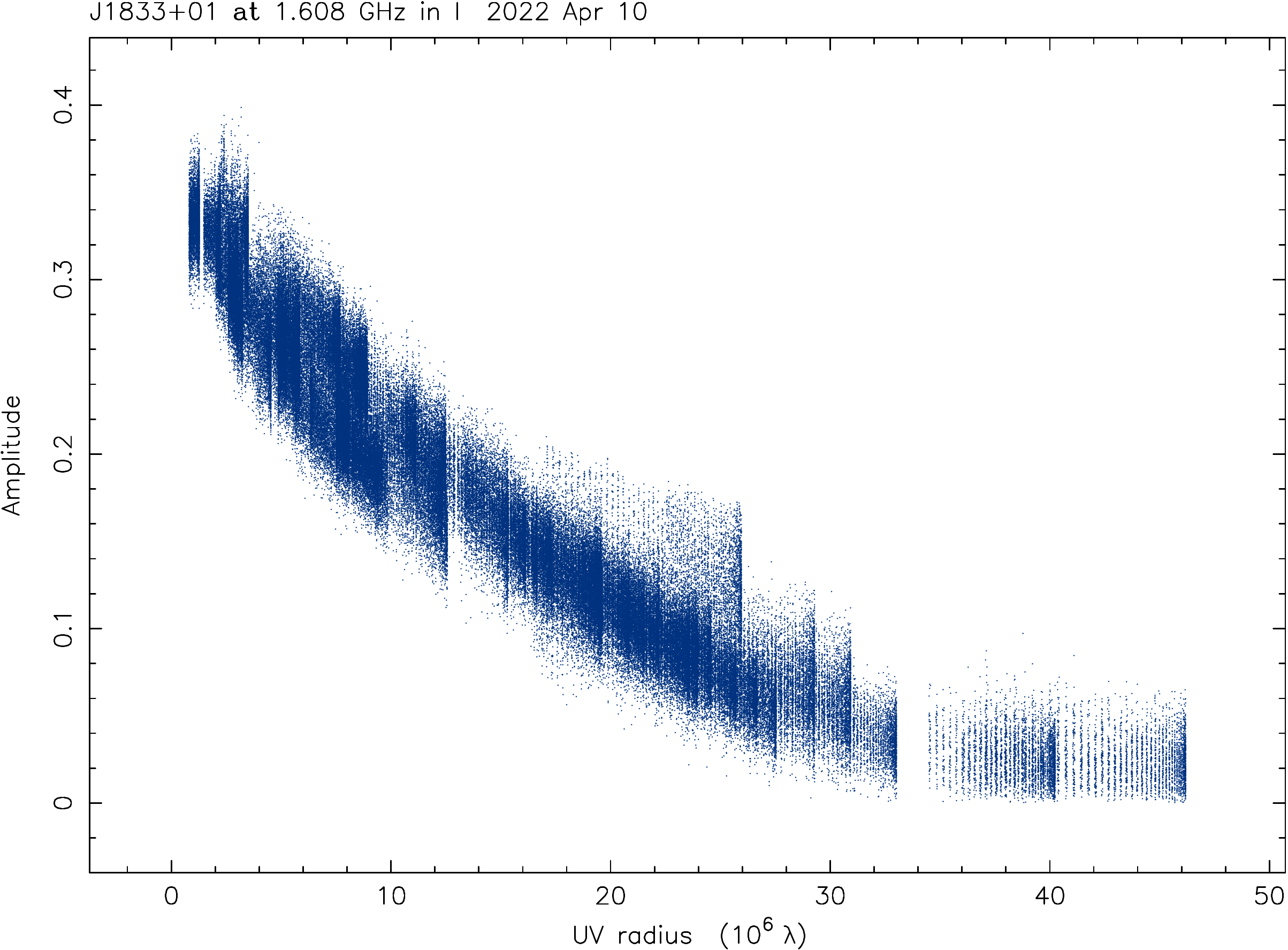}
\caption{Correlated flux density in Jy as a function of baseline projection $r_{uv}$ for the phase calibrator J1833+0115 at L-band.}
\label{fa:radplot-l-phas-cal}
\end{figure*}

In \autoref{fa:radplots_c}, \autoref{fa:radplots_l} and \autoref{fa:radplots_s}, we show amplitudes of the visibility function for the sources detected at C, L and S-bands, respectively. We note that the maximum correlated flux density detected by VLBA at short baseline spacings on average decreases with increasing source number, which is assigned (at C-band) according to the integrated VLA flux density measured in the GLOSTAR survey \citep{Medina19} at 5.8~GHz, i.e., higher number for a weaker source. \autoref{fa:radplot-l-phas-cal} presents the visibility function of the phase calibrator J1833+0115 at L-band, without averaging over time and IFs.

\begin{figure*}
\centering
\includegraphics[width=0.7\linewidth]{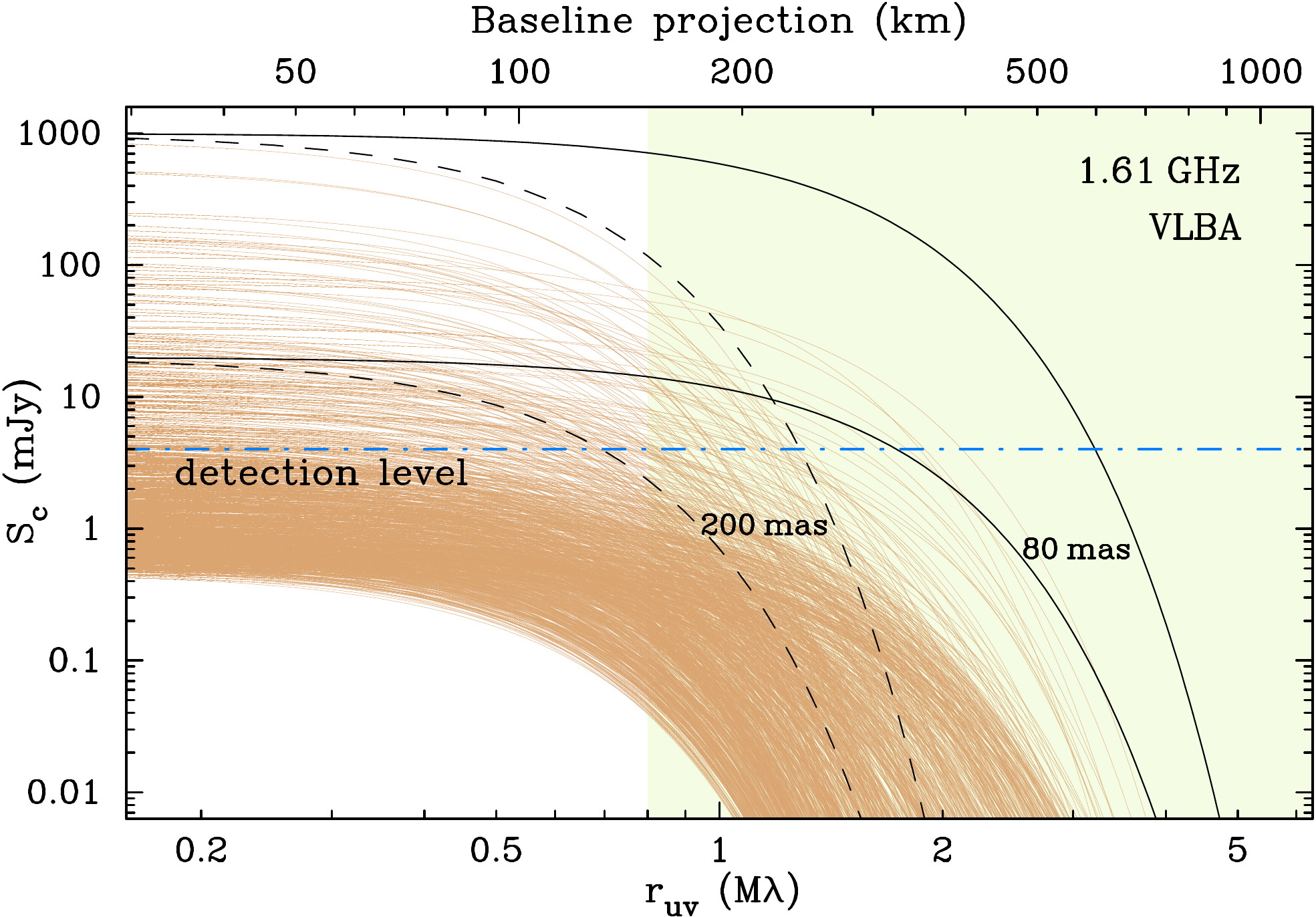}
\caption{Amplitude of visibility function expected for all observed targets at L-band (orange lines) based on a Gaussian source size $\theta(b)$ from the fit in \autoref{f:scat_size_vs_b} and total flux densities obtained within the GLOSTAR project \citep{Dzib23}. Solid and dashed black lines representing a Gaussian source of a size of 80 and 200~mas, respectively, are shown for guidance for two cases, a bright 1~Jy source and a weaker 20~mJy object. Shaded area represents the baseline range provided by the VLBA. The horizontal line denotes the approximate detection level.}
\label{fa:vis_func_L_all_sources}
\end{figure*}

Amplitude of simulated visibility functions expected for all the target sources observed at L-band are shown in \autoref{fa:vis_func_L_all_sources}. These are based on Gaussian source sizes $\theta(b)$ derived from the fit in \autoref{f:scat_size_vs_b} and the total flux densities obtained within the GLOSTAR project \citep{Dzib23}.

\section{Observational details}
\label{sec:obs_det}

Frequency setup for all the BB436 VLBA sessions and their dates are listed in \autoref{ta:freqs} and \autoref{ta:sessions}, respectively.

\begin{table*}
\centering
\caption{Frequency setup for the BB436 VLBA experiment.}
\label{ta:freqs}
\begin{tabular}{c c c c c c c c c}
\hline\hline\noalign{\smallskip}
 Band & \multicolumn{8}{c}{Frequency channels} \\
      &    IF1 &   IF2 &   IF3 &   IF4 &   IF5 &   IF6 &   IF7 &   IF8 \\
      &  (GHz) & (GHz) & (GHz) & (GHz) & (GHz) & (GHz) & (GHz) & (GHz) \\
\hline\noalign{\smallskip}
   L  &  1.392 & 1.424 & 1.456 & 1.488 & 1.648 & 1.744 & 1.776 & 1.808 \\
   S  &  2.188 & 2.220 & 2.252 & 2.284 & 2.316 & 2.348 & 2.380 & 2.412 \\
   C  &  4.612 & 4.740 & 4.868 & 4.996 &\ldots &\ldots &\ldots &\ldots \\
\hline
\end{tabular}
\end{table*}

\begin{table*}
\centering
\caption{Dates of the BB436 VLBA sessions.}
\label{ta:sessions}
\begin{tabular}{c c}
\hline\hline\noalign{\smallskip}
          Date & Band \\
  (yyyy--mm--dd) &      \\
\hline\noalign{\smallskip}
 2022--04--10 & L \\
 2022--04--17 & L \\
 2022--05--16 & L \\
 2022--06--07 & L \\
 2022--06--08 & L \\
 2022--06--13 & L \\
 \hline
 2023--05--15 & S \\
 \hline
 2023--06--12 & C \\
 2023--06--14 & C \\
 2023--06--21 & C \\
 2023--07--04 & C \\
\hline
\end{tabular}
\end{table*}

\bsp
\label{lastpage}
\end{document}